\documentclass[structabstract]{aa}
\usepackage{graphicx}
\usepackage{txfonts}
\usepackage[authoryear]{natbib}
\usepackage{longtable,lscape}

\bibpunct{(}{)}{;}{a}{}{,}
\defcitealias{Rampazzo05}{Paper~I}
\defcitealias{Annibali06}{Paper~II}
\defcitealias{Annibali07}{Paper~III}
\defcitealias{Annibali10a}{Paper~IV}
\defcitealias{Marino10}{Paper~V}

\begin{document}
\title{Nearby early--type galaxies with ionized gas VI. The {\it Spitzer}-IRS view}

\subtitle{Basic data set analysis and empirical spectral classification\thanks{Based on  {\it Spitzer}
observations: Cycle3 ID 30256 PI R. Rampazzo}}

\author{P. Panuzzo\inst{1}, R. Rampazzo \inst{2}, A. Bressan \inst{2,3,4}, O. Vega\inst{4},  
F. Annibali \inst{2,5}, L. M. Buson\inst{2}, M. S. Clemens \inst{2}, W. W. Zeilinger \inst{6}}
\institute{
CEA, Laboratoire AIM, Irfu/SAp, F-91191 Gif-sur-Yvette, France\\
\email{pasquale.panuzzo@cea.fr}
\and
INAF - Osservatorio Astronomico di Padova, Vicolo dell'Osservatorio 5, I-35122 Padova, Italy\\
\email{roberto.rampazzo@oapd.inaf.it, alessandro.bressan@oapd.inaf.it, francesca.annibali@oapd.inaf.it,
lucio.buson@oapd.inaf.it, marcel.clemens@oapd.inaf.it}
\and
SISSA, Via Beirut 4, I-34014 Trieste, Italy
\and
INAOE, Luis Enrique Erro 1, 72840, Tonantzintla, Puebla, Mexico\\
\email{ovega@inaoep.mx}
\and
STSci, 3700 San Martin Drive
Baltimore, MD 21218, USA
\and
Institut f\"ur Astronomie der Universit\"at  Wien, T\"urkenschanzstra{\ss}e 17, A-1180 Wien, Austria\\
\email{werner.zeilinger@univie.ac.at}
}

\date{Received ; accepted}
\authorrunning{Panuzzo et al.}


\abstract
{A large fraction of early-type galaxies (ETGs) shows emission lines
in their optical spectra, mostly with LINER characteristics. 
Despite the number of studies, the nature of the ionization 
mechanisms is still debated. Many ETGs also show several signs of
rejuvenation episodes.}
{We aim to investigate the ionization mechanisms and the physical processes
of a sample of ETGs using mid-infrared spectra.}
{We present here low resolution  {\it Spitzer}-IRS  spectra 
of 40 ETGs, 18 of which from our proposed Cycle~3 observations, selected from 
a sample of 65 ETGs showing emission lines in their optical spectra.
We homogeneously extract the mid-infrared (MIR) spectra, and after the proper subtraction of
a ``passive'' ETG template, we derive the intensity of the ionic and
molecular lines and of the polycyclic aromatic hydrocarbon (PAH) emission features. We
use MIR diagnostic diagrams to investigate the powering mechanisms of the
ionized gas.}
{The mid-infrared spectra of early-type galaxies show a variety of spectral
characteristics. We empirically sub-divide the sample into five classes of 
spectra with common characteristics. Class-0, accounting for 20\% of the 
sample, are purely passive ETGs with neither emission lines nor PAH 
features. Class-1 show emission lines but no PAH features, and account for
17.5\% of the sample. Class-2, in which 50\% of the ETGs are found, as well 
as having emission lines, show PAH features with unusual ratios, 
e.g. $7.7\;\rm \mu m/11.3\;\rm \mu m \leq 2.3$. Class-3 objects 
(7.5\% of the sample) have emission 
lines and PAH features with ratios typical of star-forming galaxies. Class-4, 
containing only 5\% of the ETGs, is dominated by a hot dust continuum. 
The diagnostic diagram [\ion{Ne}{III}]15.55$\mu$m/[\ion{Ne}{II}]12.8$\mu$m vs. 
[\ion{S}{III}]33.48$\mu$m/[\ion{Si}{II}]34.82$\mu$m, is used to 
investigate the different mechanisms ionizing the gas. 
According to the above diagram most of our ETGs contain gas ionized via 
either AGN-like or shock phenomena, or both.}
{Most of the spectra in the present sample are classified as
LINERs in the optical window. The proposed MIR spectral classes
show unambiguously the manifold of the physical processes and 
ionization mechanisms, from star formation, low level AGN
activity, to shocks (H$_2$), present in LINER nuclei.}

\keywords{Galaxies: elliptical and lenticular, cD -- Galaxies:
fundamental parameters -- Galaxies: evolution -- Galaxies: ISM}

\maketitle
%

\section{Introduction}
 
The view that early-type galaxies (ETGs) are inert stellar systems,
essentially devoid of gas and dust, has radically changed over the last
$\sim$ 20 years, since the increase in instrumental sensitivity across the
electromagnetic spectrum has revealed the presence of a complex
interstellar medium (ISM). Nebular emission lines are commonly found in
the inner regions of ETGs: in the optical, the documented detection
fractions for ``unbiased'' samples are 55\%--60\% \citep{Phillips86},
72\% (ellipticals, E) --85\% (lenticulars, S0)
\citep{Macchetto96}, 66\% (E) --83\% (S0) \citep{Sarzi06}, and 52\% 
\citep{Yan06}.
By means of optical emission line ratios \citep[e.g. 
\protect{[\ion{O}{III}]$\lambda$5007/H$\beta$ and
[NII]$\lambda$6584/H$\alpha$,}][]{Baldwin81}, it has also been shown
\citep[e.g.][]{Phillips86,Goudfrooij99,Kewley06,Annibali10a}
that the emission in
ETGs is ``indistinguishable'' from that of low-ionization nuclear
emission-line regions \citep[LINERs,][]{Heckman80}.

Despite the large number of studies, the excitation mechanism in LINERs is
still highly debated. Low accretion-rate AGNs are good candidates because
they are capable of reproducing the observed optical emission line ratios. This
mechanism is supported by the presence of compact X-ray and/or 
nuclear radio sources \citep[e.g.][]{Gonzalez09}, UV and X-ray
variability \citep[e.g.][]{Maoz05,Pian10},
and broad emission lines in the optical spectra, besides the fact that massive
black holes (BHs) appear to be a generic component of galaxies with a
bulge \citep[e.g.][]{Kormendy04}. On the other hand, evidence is
growing for a deficit of ionizing photons from weak AGNs
\citep[e.g.][]{Eracleous10}, suggesting that more than one excitation mechanism
may operate in LINERs. Fast shocks \citep{Koski76,Heckman80,Dopita95,Allen08}
and photoionization by old post-asymptotic giant
branch (PAGB) stars \citep{Trinchieri91,Binette94,Stasinska08} have been
proposed as alternative mechanisms. However, we have shown in
\citet[][Paper IV hereafter]{Annibali10a}
that PAGB stars can account for the ionizing photon
budget only in the weakest LINERs, or in off-nuclear regions. Indeed, for
almost 80\% of our sample, either low accretion-rate AGNs and/or fast
shocks in a relatively gas poor environment are needed.

The high sensitivity of the Infrared Spectrograph (IRS) on board the
{\it Spitzer} Space Telescope opened the mid-infrared (MIR) window to the
study of ETGs. This spectral window is particularly rich in information
on the ISM, excitation mechanisms and galaxy evolution.

\citet{Bressan06} showed that the MIR spectra of the majority of the massive ETGs in clusters
show neither Polycyclic Aromatic Hydrocarbons (PAHs) features nor emission lines, 
and exhibit only the broad silicate emission feature around 10$\mu$m arising
from circumstellar dust around
oxygen-rich AGB stars, superimposed on a stellar photospheric continuum.
ETGs with such MIR spectra, with either un-excited, or likely absent 
\citep[see][]{Clemens10}, 
ISM, represent the class of passively evolving ETGs and can be considered
the fossil record of galaxy evolution.

Another interesting result from {\it Spitzer} is that several ETGs exhibit 
PAH emission features with ``unusual'' ratios \citep[e.g.][]{Kaneda05,Kaneda08}: 
usually strong emission features at 
6.2, 7.7, and 8.6~$\mu$m are weak in contrast to 
prominent features at 12.7 and 11.3~$\mu$m.
This may reflect peculiar physical conditions in the ISM, where the 
PAH emission is not powered by star formation activity.

MIR diagnostic diagrams were presented by \citet{Sturm06}, and
\citet{Dale06,Dale09} to study ISM excitation mechanisms.
\citet{Sturm06} compared the MIR properties 
(SED, PAHs, and emission line ratios) of IR-luminous and IR-faint LINERs. 
They found that IR-luminous LINERs have MIR SEDs that are 
similar to those of starburst galaxies, and are situated in different 
regions of the diagnostic diagrams than IR-faint LINERs. 
From the presence of strong [\ion{O}{IV}]25.9$\mu$m emission, indicative 
of highly ionized gas, they suggested a low-luminosity AGN powering 
source in 90\%  in both subsamples  \citep[see also][]{Rupke07}.
\citet{Dale06,Dale09} devised MIR 
diagnostic diagrams based on both ionic lines and PAH features, which 
are particularly suitable in distinguishing between AGN and starburst excitation.

The present paper is part of a series dedicated to the study of nearby
ETGs with emission lines, selected from the original optical 
sample of 65 galaxies in \citet[][Paper I hereafter]{Rampazzo05} 
and \citet[][Paper II hereafter]{Annibali06}, and
it presents the MIR properties of 40 ETGs obtained from 
low-resolution {\it Spitzer}-IRS spectra.
We analyze and subdivide spectra into empirical classes, according to their properties,
and propose these latter represent possible phases of the nuclear
evolution induced by an accretion event.

The paper is organized as follows. Section~\ref{sec:sample} presents an overview 
of the sample, summarizing some of the results obtained from the optical.
In Sect.~\ref{sec:data}, we describe the procedure adopted to
extract the spectra. In Sect.~\ref{sec:analysis} we analyze the MIR spectra and 
propose an empirical classification of ETG spectra based on their
properties and on the measurement of the ionic/molecular 
emission lines and PAH features. In Sect.~\ref{sec:diagnostic}, we investigate MIR diagnostic diagrams. 
In Sect.~\ref{sec:discussion} we discuss how MIR spectral classes can be interpreted as characteristic
of different phases in an evolutionary scenario.
Our conclusions are presented in Sect.~\ref{sec:conclusions}.

\begin{table*}
\caption{The sample overview}\label{tab1}
\centering
\begin{tabular}{llccccrcc}
\hline
\hline
{ident} & {RSA} & {RC3} &  {V$_{hel}$}      &  {$\rho_{xyz}$}       & $\sigma_{r_e/8}$ & Age  & Z & [$\alpha$/Fe] \\
              & {}          &  {}        &  {km~s$^{-1}$} &  {Gal. Mpc$^{-3}$} &  {km~s$^{-1}$}     & Gyr  &    & \\
\hline 
NGC~1052 & E3/S0       & E4        & 1510 & 0.49 & 215 & 14.5 $\pm$ 4.2 & 0.032 $\pm$      0.007     &    0.34    $\pm$     0.05  \\
NGC~1209 & E6          & E6:       & 2600 & 0.13 & 240 &  4.8 $\pm$ 0.9 & 0.051 $\pm$      0.012     &    0.14    $\pm$     0.02 \\
NGC~1297 & S02/3(0)    & SAB0 pec: & 1586 & 0.71 & 115 & 15.5 $\pm$ 1.2 & 0.012 $\pm$      0.001     &    0.29    $\pm$ 0.04 \\
NGC~1366 & E7/S01(7)   & S0 sp     & 1231 & 0.16 & 120 &  5.9 $\pm$ 1.0 & 0.024 $\pm$      0.004     &    0.08    $\pm$     0.03 \\
NGC~1389 & S01(5)/SB01 & SAB(s)0-: &  912 & 1.50 & 139 &  4.5 $\pm$ 0.6 & 0.032 $\pm$      0.005     &    0.08    $\pm$     0.02 \\
NGC~1407 & E0/S01(0)   & E0        & 1779 & 0.42 & 286 &  8.8 $\pm$ 1.5 & 0.033 $\pm$      0.005     &    0.32    $\pm$     0.03 \\
NGC~1426 & E4          & E4        & 1443 & 0.66 & 162 &  9.0 $\pm$ 2.5 & 0.024 $\pm$      0.005     &    0.07    $\pm$     0.05  \\
NGC~1453 & E0          & E2        & 3886 & ...  & 289 &  9.1 $\pm$ 2.8 & 0.034 $\pm$      0.009     &    0.22    $\pm$     0.05  \\
NGC~1533 & SB02(2)/SBa & SB0-      &  790 & 0.89 & 174 & 11.9 $\pm$ 6.9 & 0.023 $\pm$      0.020     &    0.21    $\pm$     0.10 \\
NGC~1553 & S01/2(5)pec & SA(r)0    & 1080 & 0.97 & 180 &  4.8 $\pm$ 0.7 & 0.031 $\pm$      0.004     &    0.10    $\pm$     0.02  \\
NGC~2974 & E4          & E4        & 1919 & 0.26 & 220 & 13.9 $\pm$ 3.6 & 0.021 $\pm$      0.005     &    0.23    $\pm$     0.06 \\
NGC~3258 & E1          & E1        & 2792 & 0.72 & 271 &  4.5 $\pm$ 0.8 & 0.047 $\pm$      0.013     &    0.21    $\pm$     0.03 \\
NGC~3268 & E2          & E2        & 2800 & 0.69 & 227 &  9.8 $\pm$ 1.7 & 0.023 $\pm$      0.004     &    0.34    $\pm$     0.04  \\
NGC~3557 & E3          & E3        & 3088 & 0.28 & 265 &  5.8 $\pm$ 0.8 & 0.034 $\pm$      0.004     &    0.17    $\pm$     0.02  \\
NGC~3818 & E5          & E5        & 1701 & 0.20 & 191 &  8.8 $\pm$ 1.2 & 0.024 $\pm$      0.003     &    0.25    $\pm$     0.03  \\
NGC~3962 & E1          & E1        & 1818 & 0.32 & 225 & 10.0 $\pm$ 1.2 & 0.024 $\pm$      0.003     &    0.22    $\pm$     0.03 \\
NGC~4374 & E1          & E1        & 1060 & 3.99 & 282 &  9.8 $\pm$ 3.4 & 0.025 $\pm$      0.010     &    0.24    $\pm$     0.08 \\
NGC~4552 & S01(0)      & E         &  340 & 2.97 & 264 &  6.0 $\pm$ 1.4 & 0.043 $\pm$      0.012     &    0.21    $\pm$     0.03 \\
NGC~4636 & E0/S01(6)   & E0-1      &  938 & 1.33 & 209 & 13.5 $\pm$ 3.6 & 0.023 $\pm$      0.006     &    0.29    $\pm$   0.06 \\
NGC~4696 &(E3)         & E+1 pec   & 2958 & 0.00 & 254 & 16.0 $\pm$ 4.5 & 0.014 $\pm$      0.004     &    0.30    $\pm$     0.10 \\
NGC~4697 & E6          & E6        & 1241 & 0.60 & 174 & 10.0 $\pm$ 1.4 & 0.016 $\pm$      0.002     &    0.14    $\pm$     0.04  \\
NGC~5011 & E2          & E1-2      & 3159 & 0.27 & 249 &  7.2 $\pm$ 1.9 & 0.025 $\pm$      0.008     &    0.25    $\pm$     0.06  \\
NGC~5044 & E0          & E0        & 2782 & 0.38 & 239 & 14.2 $\pm$ 10. & 0.015 $\pm$      0.022     &    0.34    $\pm$     0.17 \\
NGC~5077 & S01/2(4)    & E3+       & 2806 & 0.23 & 260 & 15.0 $\pm$ 4.6 & 0.024 $\pm$      0.007     &    0.18    $\pm$     0.06   \\
NGC~5090 & E2          & E2        & 3421 &  ... & 269 & 10.0 $\pm$ 1.7 & 0.028 $\pm$      0.005     &    0.26    $\pm$     0.04  \\
NGC~5638 & E1          & E1        & 1676 & 0.79 & 165 &  9.1 $\pm$ 2.3 & 0.024 $\pm$      0.008     &    0.24    $\pm$     0.05 \\
NGC~5812 & E0          & E0        & 1970 & 0.19 & 200 &  8.5 $\pm$ 2.1 & 0.027 $\pm$      0.008     &    0.22    $\pm$     0.05 \\
NGC~5813 & E1          & E1-2      & 1972 & 0.88 & 239 & 11.7 $\pm$ 1.6 & 0.018 $\pm$      0.002     &    0.26    $\pm$     0.04 \\
NGC~5831 & E4          & E3        & 1656 & 0.83 & 164 &  8.8 $\pm$ 3.5 & 0.016 $\pm$      0.011     &    0.21    $\pm$     0.09\\
NGC~5846 & S01(0)      & E0+       & 1714 & 0.84 & 250 &  8.4 $\pm$ 1.3 & 0.033 $\pm$      0.005     &    0.25    $\pm$     0.03 \\
NGC~5898 & S02/3(0)    & E0        & 2122 & 0.23 & 220 &  7.7 $\pm$ 1.3 & 0.030 $\pm$      0.004     &    0.10    $\pm$     0.03 \\
NGC~6868 & E3/S02/3(3) & E2        & 2854 & 0.47 & 277 &  9.2 $\pm$ 1.8 & 0.033 $\pm$      0.006     &    0.19    $\pm$     0.03 \\
NGC~7079 & SBa         & SB(s)0    & 2684 & 0.19 & 155 &  6.7 $\pm$ 1.1 & 0.016 $\pm$      0.003     &    0.21    $\pm$     0.05    \\
NGC~7192 & S02(0)      & E+:       & 2978 & 0.28 & 257 &  5.7 $\pm$ 2.0 & 0.039 $\pm$      0.015     &    0.09    $\pm$     0.05 \\
NGC~7332 & S02/3(8)    & S0 pec sp & 1172 & 0.12 & 136 &  3.7 $\pm$ 0.4 & 0.019 $\pm$      0.002     &    0.10    $\pm$     0.03 \\
IC~1459  & E4          & E         & 1802 & 0.28 & 311 &  8.0 $\pm$ 2.2 & 0.042 $\pm$      0.009     &    0.25    $\pm$     0.04  \\
IC~2006  & E1          & E         & 1382 & 0.12 & 122 &  8.1 $\pm$ 1.7 & 0.027 $\pm$      0.005     &    0.12    $\pm$     0.04  \\
IC~3370  & E2 pec      & E2+       & 2930 & 0.20 & 202 &  5.6 $\pm$ 0.9 & 0.022 $\pm$      0.004     &    0.17    $\pm$     0.04 \\
IC~4296  & E0          & E         & 3737 & ...  & 340 &  5.2 $\pm$ 1.0 & 0.044 $\pm$      0.008     &    0.25    $\pm$     0.02\\
IC~5063  & S03(3)pec/Sa& SA(s)0+:  & 3402 & ...  & 160 &  ... & ... & ... \\
\hline
\end{tabular}
\tablefoot{
See Sect.~\ref{sec:sample}  for a detailed explanation of single columns.
The age, metallicity and the $\alpha$-enhancement obtained from the Lick line-strength
index analysis, are obtained from \citetalias{Annibali07}.}
\end{table*}

\begin{table*}
\caption {The {\it Spitzer}--IRS observations.}
\centering
\begin{tabular}{llrrrrr}
\hline
\hline
{Ident} &  {PI}  &  {ID} & {SL1}                     &   {SL2}      &  {LL2}                    &  {LL1}  \\
{}          & {}        &   {}     &  [s$\times$Cycle] &  [s$\times$Cycle] & [s$\times$Cycle]  & [s$\times$Cycle] \\
\hline
\object{NGC~1052} & Kaneda & 30483 &60$\times$2 & 60$\times$2 & 30$\times$2& 30$\times$2\\
\object{NGC~1209} & Rampazzo & 30256 &  60$\times$6  & 60$\times$6  & 120$\times$16& 120$\times$8\\
\object{NGC~1297} & Rampazzo & 30256 & 60$\times$19  & 60$\times$19  & 120$\times$14& 120$\times$8\\
\object{NGC~1366} & Rampazzo & 30256 & 60$\times$11  & 60$\times$11  & 120$\times$14& 120$\times$8\\
\object{NGC~1389} & Rampazzo & 30256 &60$\times$9  & 60$\times$9  & 120$\times$14& 120$\times$8\\
\object{NGC~1407} & Kaneda &3619/30483 &60$\times$2 & 60$\times$2 & 30$\times$2& 30$\times$2\\
\object{NGC~1426} & Rampazzo & 30256 & 60$\times$12  & 60$\times$12  & 120$\times$14& 120$\times$8\\
\object{NGC~1453} & Bregman   &  3535         & 14$\times$8 &  14$\times$8  &   30$\times$6 & ... \\
\object{NGC~1533} & Rampazzo  & 30256 & 60$\times$3  & 60$\times$3  & 120$\times$5& 120$\times$3\\
\object{NGC~1553} & Rampazzo & 30256 & 60$\times$3  & 60$\times$3  & 120$\times$3& 120$\times$3\\
\object{NGC~2974} & Kaneda  &3619/30483 &60$\times$2 & 60$\times$2 & 30$\times$3& 30$\times$3\\
\object{NGC~3258} & Rampazzo &30256 & 60$\times$8  & 60$\times$8  & 120$\times$14& 120$\times$8\\
\object{NGC~3268} & Rampazzo &30256  & 60$\times$9  & 60$\times$9  & 120$\times$14& 120$\times$8\\
\object{NGC~3557} & Kaneda  & 30483 &   60$\times$3  & 60$\times$3  & 30$\times$3& 30$\times$3\\  
\object{NGC~3818} & Rampazzo & 30256 &60$\times$19  & 60$\times$19  & 120$\times$14& 120$\times$8\\
\object{NGC~3962} & Kaneda &3619/30483 &60$\times$2 & 60$\times$2 & 30$\times$3& 30$\times$3\\
\object{NGC~4374} & Rieke & 82 & 60$\times$4  & 60$\times$4  & 120$\times$4&120$\times$4\\
\object{NGC~4552} & Bregman & 3535 &14$\times$8  & 14$\times$8  & 30$\times$6& ...\\
\object{NGC~4636} & Bressan &3419 &  60$\times$3  & 60$\times$3  & 120$\times$5 & ...\\
\object{NGC~4696} & Kaneda &3619/30483 &60$\times$2 & 60$\times$2 & 30$\times$3& 30$\times$3\\
\object{NGC~4697} & Bregman &  3535 &14$\times$8  & 14$\times$8  & 30$\times$6& ...\\
\object{NGC~5011} & Rampazzo &30256 &60$\times$6  & 60$\times$6  & 120$\times$12& 120$\times$8\\
\object{NGC~5044} & Rampazzo & 30256 & 19 & 19 & 14 & 8\\
\object{NGC~5077} & Rampazzo & 30256 &60$\times$12  & 60$\times$12  & 120$\times$14& 120$\times$8\\
\object{NGC~5090} & Kaneda &30483& 60$\times$4 & 60$\times$4 & 30$\times$3& 30$\times$3\\
\object{NGC~5638} & Bregman &  3535 &14$\times$8  & 14$\times$8  & 30$\times$6& ...\\
\object{NGC~5812} & Bregman & 3535 &60$\times$6  & 60$\times$6  & 120$\times$12& 120$\times$8\\
\object{NGC~5813} & Bregman & 3535 & 14$\times$8  & 14$\times$8  & 30$\times$6& ...\\
\object{NGC~5831} & Bregman &  3535 & 14$\times$8  & 14$\times$8  & 30$\times$6& ...\\
\object{NGC~5846} & Bregman &3535 & 14$\times$8  & 14$\times$8  & 30$\times$6& ...\\
\object{NGC~5898} & Rampazzo & 30256 &60$\times$11  & 60$\times$11  & 120$\times$14& 120$\times$8\\
\object{NGC~6868} & Rampazzo &30256 &60$\times$6  &  60$\times$6 &  120$\times$13 &120$\times$8\\
\object{NGC~7079} & Rampazzo & 30256 & 60$\times$19 &60$\times$19  & 120$\times$14  & 120$\times$8\\
\object{NGC~7192} & Rampazzo&  30256 & 60$\times$12 &60$\times$12  & 120$\times$14  & 120$\times$8\\
\object{NGC~7332} & Rampazzo & 30256 &60$\times$7 &60$\times$7  & 120$\times$14  & 120$\times$8\\
\object{IC~1459}  & Kaneda &30483  & 60$\times$3 & 60$\times$3 & 30$\times$2& 30$\times$2\\
\object{IC~2006}  & Bregman             &  3535         & 14$\times$8 &  14$\times$8  &   30$\times$6 & ...\\
\object{IC~3370}  & Kaneda & 3619/30483 &60$\times$2 & 60$\times$2 & 30$\times$3& 30$\times$3\\
\object{IC~4296}  & Antonucci & 20525 & 240$\times$2  & 240$\times$2  & 120$\times$3& 120$\times$3\\
\object{IC~5063}  & Gorjan & 30572 & 14$\times$2  & 14$\times$2  & 30$\times$1& 30$\times$1\\
\hline
\end{tabular}
\label{tab2}
\end{table*}


\begin{figure*}[!ht]
\centering
\includegraphics[width=0.95\textwidth]{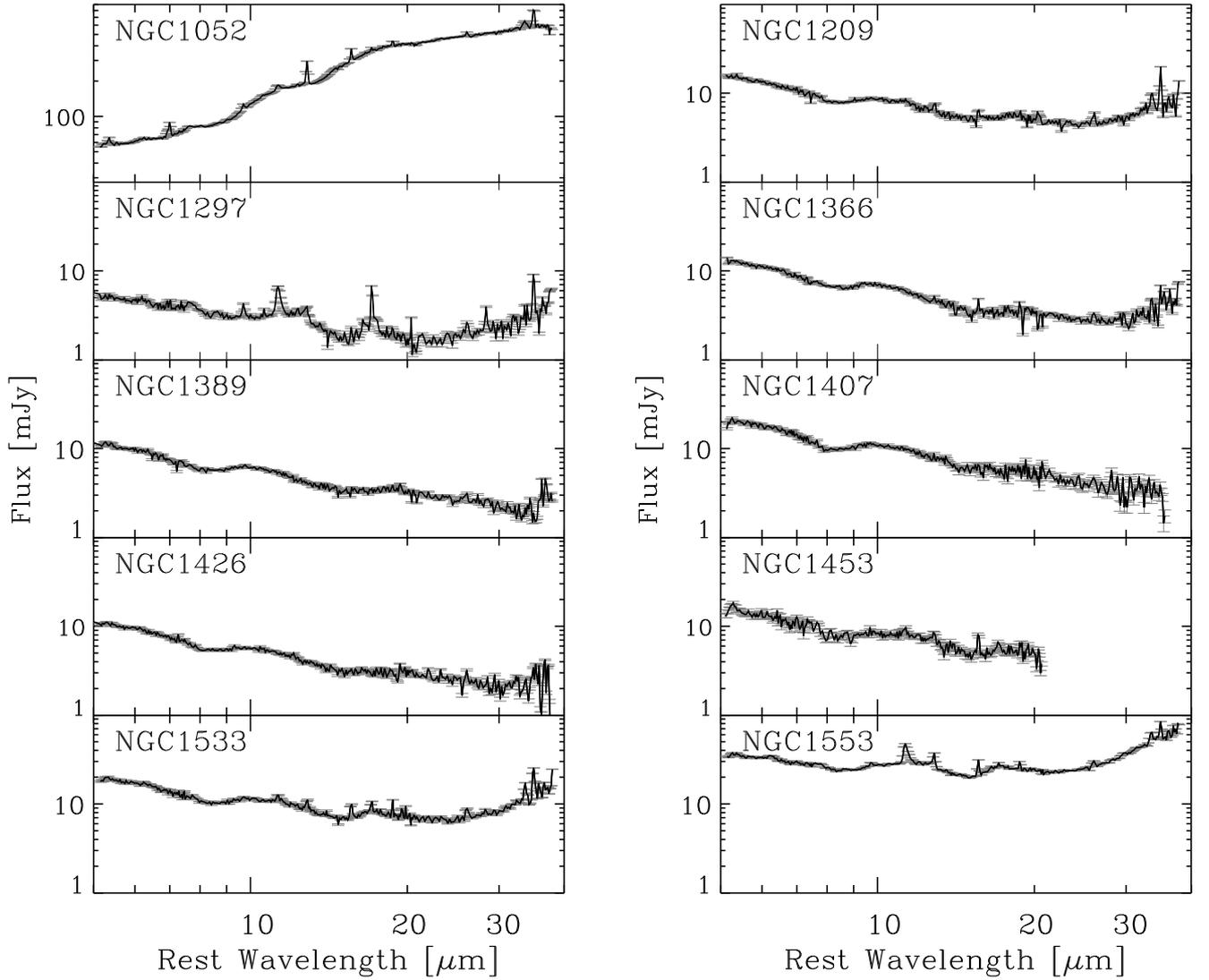}
\caption{ETG MIR spectra. Flux vs rest wavelength as obtained from
{\it Spitzer}-IRS low resolution modules. The LL modules have been
scaled to SL fluxes.
}
\label{Fig1}%
\end{figure*}
%

%
\begin{figure*}
\centering
\includegraphics[width=0.95\textwidth]{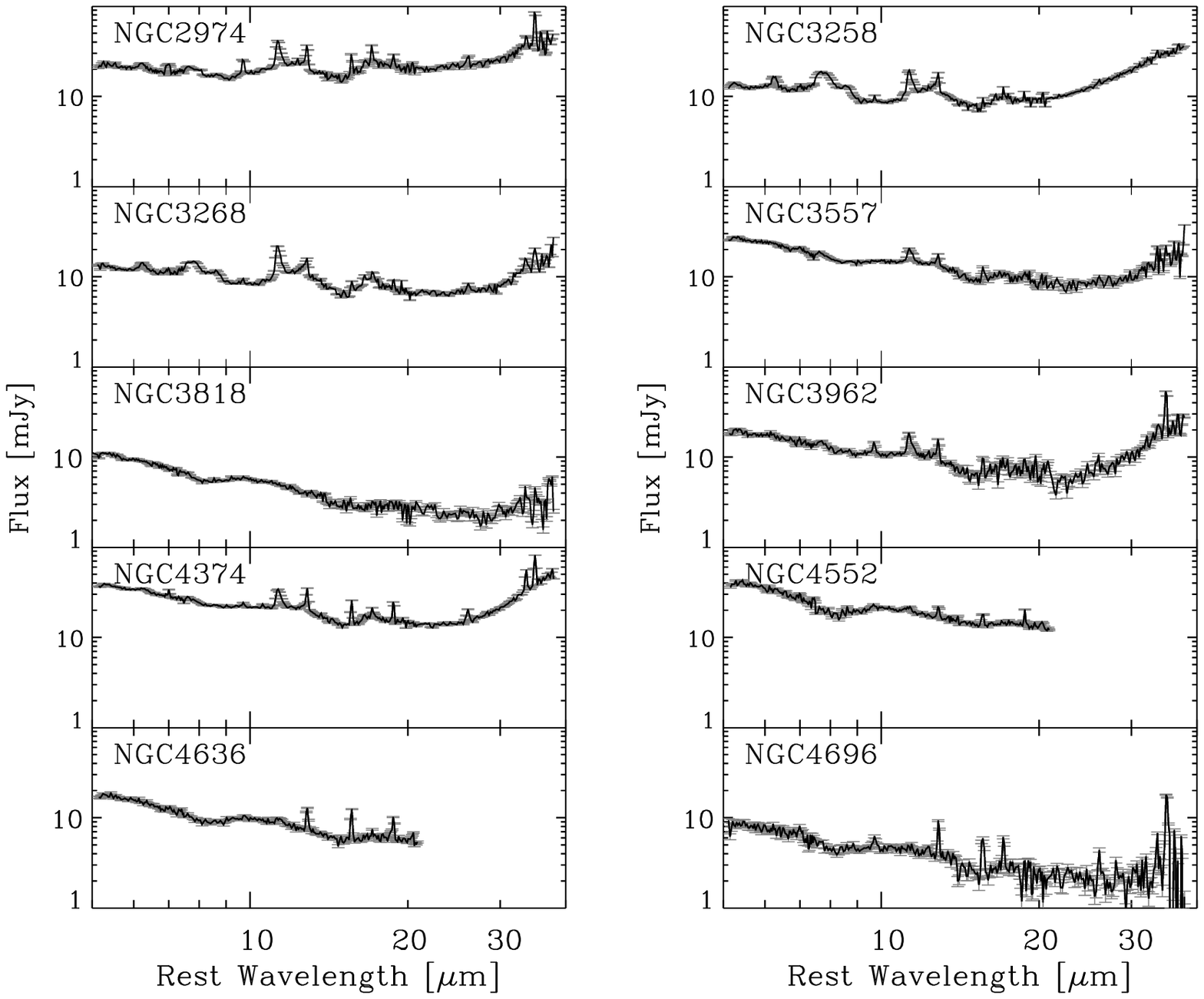}
\addtocounter{figure}{-1}
\caption{Continue.}
\end{figure*}
%

%
\begin{figure*}
\centering
\includegraphics[width=0.95\textwidth]{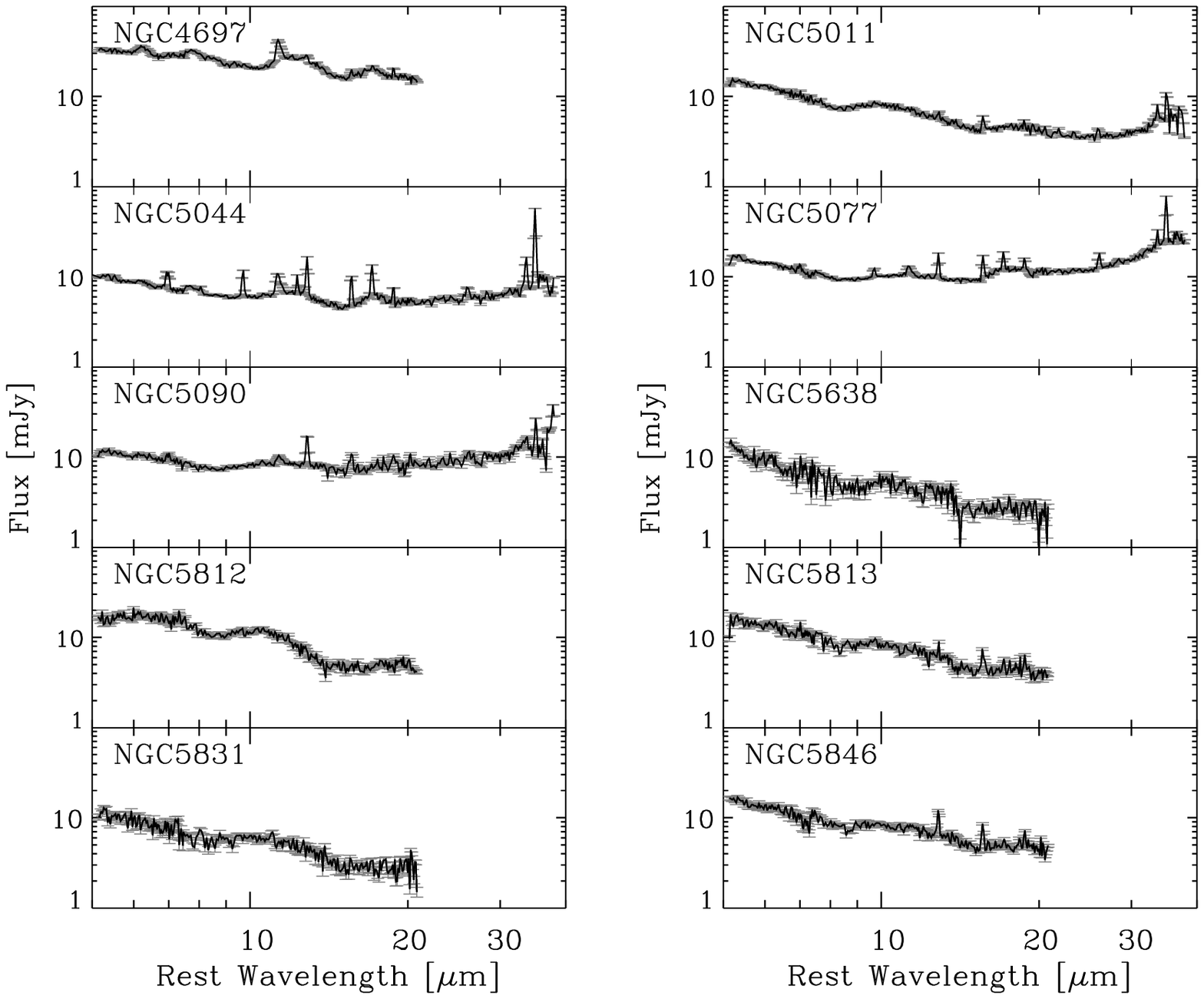}
\addtocounter{figure}{-1}
\caption{Continue.}
\end{figure*}
%

\begin{figure*}
\centering
\includegraphics[width=0.95\textwidth]{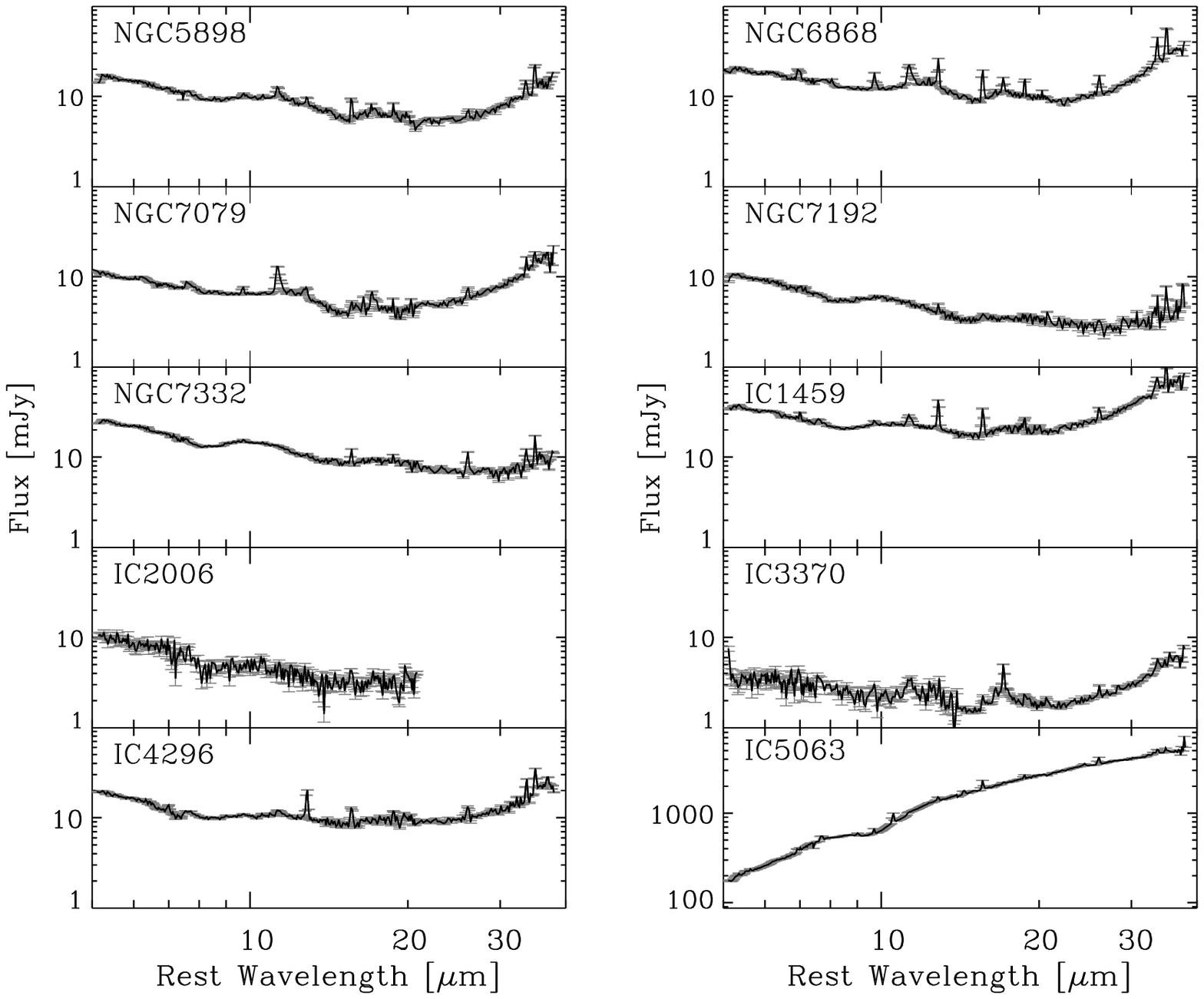}
\addtocounter{figure}{-1}
\caption{Continue.  For IC~3370 \citep[see][]{Kaneda08},
the SL observations missed the center of the galaxy by 10\arcsec, which
caused significant reduction in the S/N ratio of the SL spectrum.}
\end{figure*}
%

\section{Sample overview}
\label{sec:sample}

The 40 galaxies presented in this paper are a subset of the original sample 
of 65 ETGs described in \citet[][Paper I]{Rampazzo05} and \citet[][Paper II]{Annibali06} 
(hereafter the R05$+$A06 sample).

The R05$+$A06 sample was selected from a compilation of ETGs exhibiting 
evidence of an ISM in at least one of the following bands: IRAS 
100 $\mu$m, X-rays, radio, HI and CO \citep{Roberts91}. 
All galaxies belong to the \emph{Revised Shapley Ames Catalog 
of Bright Galaxies (RSA)} \citep{RSA} and 
have recessional velocities lower than $\approx$5000 km  s$^{-1}$. 
For the major fraction of the sample, the X-ray luminosities are fainter than 
$\log L_X \sim 41.88$ erg~s$^{-1}$ \citep{OSullivan01}, 
far below the luminosities of powerful AGNs.
Because of the selection criteria, the sample may be biased towards 
the presence of emission lines.

The 40 ETGs in the present sample are partly derived 
from the {\it Spitzer} archive while 18 have been observed as part of our original
Cycle~3 proposal. At the time of the proposal submission, 19 ETGs out of 65 
were already observed during Cycle~1 and Cycle~2. According to our optical data,  
some of them are among  the more ``active'' ETGs in the R05+A06 sample. 
Observing constraints apart,  our 18 targets  have been 
selected rejecting ETGs with F$_{60\mu m}$/K$_{Stot} \geq$ 0.6,
where   F$_{60\mu m}$ is the IRAS flux at 60$\mu$m and
K$_{Stot}$ the total 2MASS K$_S$-band magnitude, to include also
``less active'' ETGs in our MIR analysis.

We detail the main properties of the galaxies in the sample studied here 
in Table~\ref{tab1}.  
Column (1) gives the galaxy identification;
Cols. (2) and (3) provide the galaxy morphological classification
according to RSA \citep{RSA} and
RC3 \citep{RC3} respectively: only in a few cases
do the two catalogues disagree in the distinction between E and S0 classes;
Col. (4) provides the galaxy systemic velocity from 
NED\footnote{NASA/IPAC Extragalactic Database ({\tt NED})},  V$_{hel}$, which is lower
than $\sim$5000 km~s$^{-1}$; Col. (5) provides the richness
parameter $\rho_{xyz}$ in (galaxy) per Mpc$^{-3}$ \citep{Tully88}.
The galaxies of our sample are mainly located in low
density environments, with $\rho_{xyz}$ varying from 
$\approx$ 0.1, characteristic of very isolated galaxies, to
$\rho_{xyz}$ $\approx$ 4, which is characteristic of denser galaxy regions
 in the Virgo cluster.  For comparison, in the \citet{Tully88} catalogue, 
$\rho_{xyz}$=1.50 for NGC~1389, which is a Fornax cluster member.
In Col. (6) we report the galaxy's central velocity dispersion,
$\sigma$, derived within an aperture of 1/8 of the effective radius, r$_e$.

The sample exhibits a large age spread, with SSP-equivalent ages 
ranging from a few Gyrs to a Hubble time \citep[][Paper III hereafter]{Annibali07}.
The metallicities and [$\alpha$/Fe] ratios are supersolar.
Both the total metallicity and  [$\alpha$/Fe] exhibit a positive 
correlation with the central velocity dispersion, indicating that the 
chemical enrichment was more efficient and the duration 
of the star formation shorter in more massive galaxies.
Recent star formation episodes appear frequent in the lowest 
density environments.

A large fraction of the galaxies ($\approx$50\%) exhibit 
morphological and kinematical peculiarities
\citepalias[see Appendix A in]{Annibali10a}.
NGC~1553, NGC~2974, NGC~4552, NGC~4696, 
NGC~5846, NGC 7192, and IC~1459 present shell structures, and 
NGC~1533, NGC~2974, and IC~5063 exhibit ring/arm-like features, 
clearly visible in  far-UV imaging \citep[][Paper V hereafter]{Marino10}. 
Noteworthy features, as reported in the literature,
are provided in the on-line notes in R05$+$A06.

The optical emission line properties of the R05$+$A06 sample 
were analyzed in \citetalias{Annibali10a}. The good signal-to-noise ratio of the 
optical spectra allowed us 
to analyze the emission lines out to a galactocentric distance 
of  half of the effective radius $r_e$.
Thanks to a proper subtraction of the underlying stellar continuum, we were able to 
detect even very faint emission lines, concluding that emission is present in 
89\% of the sample. It is interesting that, despite the sample selection criteria, 
the incidence of emission is not too far from that in other 
samples of ETGs, such as that of SAURON \citep{deZeeuw02}.

From the nuclear ($r<r_e/16$) emission line ratios, the majority of the galaxies are 
classified as LINERs, while Seyferts and transition/composite 
(HII-LINERs) objects are present in minor fractions.
The nuclear (r$<r_e/16$) emission can be attributed to photoionization 
by PAGB stars alone only for $\approx$ 22\% of the LINER/composite sample.
On the other hand, we cannot exclude an important role of PAGB stellar
photoionization at larger radii. For the major fraction of the sample, the nuclear
emission is consistent with excitation caused by either a low-accretion rate AGN 
or fast shocks (200--500 km/s) in a relatively gas poor environment
(n $\lesssim100$ cm$^{-3}$), or both. 
The derived [\ion{S}{II}]$\lambda$6717/$\lambda$6731 ratios are consistent
with the low gas densities required 
by the shock models. The average nebular oxygen abundance is slightly less than solar, 
and a comparison with the results obtained from Lick indices in \citetalias{Annibali07}
reveals that it is $\approx$ 0.2 dex lower than that of the stars.
This suggests an {\em external origin for at least part of the gas}.


\onllongtabL{3}{
\begin{landscape}
\begin{longtable}{lcccccccccc}
\caption{MIR emission line intensities.}\\
\hline
\hline
Galaxy & H$_2$ 0--0 S(7) & H$_2$ 0--0 S(6) & H$_2$ 0--0 S(5) & $[$\ion{Ar}{II}] & HPf$\alpha$ & H$_2$ 0--0 S(4) &
$[$\ion{Ar}{III}] & H$_2$ 0--0 S(3) & $[$\ion{S}{IV}] & H$_2$ 0--0 S(2) \\ 
name & 5.51$\mu$m & 6.11$\mu$m & 6.91$\mu$m & 6.99$\mu$m & 7.46$\mu$m & 8.03$\mu$m & 8.99$\mu$m & 9.66$\mu$m &
 10.51$\mu$m & 12.28$\mu$m \\
\hline
NGC~1052 & 17.8 $\pm$ 3.1 & 8.7 $\pm$ 2.7 & 31.5 $\pm$ 7.8   & 81.0 $\pm$ 2.7  & ... & 6.7 $\pm$ 2.0 & 17.0 $\pm$ 2.9   & 63.3 $\pm$ 10.7 & ...               & 11.2 $\pm$ 3.9 \\
NGC~1209 & ...            & ...           & ...              & 4.8 $\pm$ 1.0    & ...           & ...           & 3.5 $\pm$ 0.9    & ...             & ...               &  1.4 $\pm$ 0.2 \\
NGC~1297 & ...            & ...           & 2.8 $\pm$ 0.3    & 3.3 $\pm$ 0.3    & ...           & 0.5 $\pm$ 0.1 & 1.2 $\pm$ 0.3    &  5.1 $\pm$ 0.5  & 0.6 $\pm$ 0.2     &  2.1 $\pm$ 0.1 \\
NGC~1366 & ...            & ...           & ...              & ...              & ...           & ...           & ...              & ...             & ...               & ...           \\
NGC~1453 & ...            & 8.5 $\pm$ 1.7 & ...              & 9.0 $\pm$ 1.1    & 3.9 $\pm$ 0.6 & ...           & ...              & 7.1 $\pm$ 1.7   & 2.2 $\pm$ 1.0     & 1.1 $\pm$ 0.2 \\
NGC~1533 & ...            & ...           & ...              & 2.2 $\pm$ 0.5    & 3.8 $\pm$ 1.2 & ...           & ...              & 3.6 $\pm$ 0.7   & ...               & 1.8 $\pm$ 0.4 \\
NGC~1553 & ...            & 1.5 $\pm$ 0.3 & 5.9 $\pm$ 0.8    & 10.6 $\pm$ 1.2   & ...           & 3.7 $\pm$ 0.4 & ...              & 10.8 $\pm$ 2.7  & 3.7 $\pm$ 0.8     & 6.2 $\pm$ 0.6 \\
NGC~2974 & 9.0 $\pm$ 2.5  & 3.8 $\pm$ 0.5 & 16.1 $\pm$ 0.8   & 18.7 $\pm$ 1.4   & ...           & 4.2 $\pm$ 0.3 & ...              & 30.9 $\pm$ 1.7  & 4.1 $\pm$ 0.6     & 13.3 $\pm$ 0.8 \\
NGC~3258 & ...            & ...           & 4.1 $\pm$ 0.3    & 8.5 $\pm$ 0.5    & ...           & 2.3 $\pm$ 0.1 & 0.8 $\pm$ 0.1    & 5.1 $\pm$ 0.8   & 0.7 $\pm$ 0.2     & 4.2 $\pm$ 0.2 \\
NGC~3268 & ...            & ...           & 6.1 $\pm$ 0.6    & 7.7 $\pm$ 0.5    & ...           & 1.6 $\pm$ 0.1 & 0.5 $\pm$ 0.1    & 4.5 $\pm$ 0.7   & 1.6 $\pm$ 0.3     & 2.4 $\pm$ 0.2 \\
NGC~3557 & ...            & ...           & ...              & 5.4 $\pm$ 0.6    & ...           & ...           & ...              & ...             & ...               & ...           \\
NGC~3962 & ...            & ...           & 3.1 $\pm$ 0.7    & 8.6 $\pm$ 1.0    & ...           & 4.5 $\pm$ 0.4 & 2.5 $\pm$ 0.7    & 16.6 $\pm$ 1.5  & 2.2 $\pm$ 0.6     & 4.8 $\pm$ 0.3 \\
NGC~4374 & ...            & ...           & 7.4 $\pm$ 1.0    & 24.4 $\pm$ 2.2   & ...           & 1.3 $\pm$ 0.1 & 6.9 $\pm$ 1.5    & 8.3 $\pm$ 2.0   & 5.2 $\pm$ 1.2     & 3.6 $\pm$ 0.5 \\
NGC~4552 & ...            & ...           & ...              & ...              & ...           & ...           & ...              & 10.4 $\pm$ 3.4  & ...               & 2.8 $\pm$ 0.9 \\
NGC~4636 & ...            & ...           & ...              & ...              & ...           & ...           & ...              & ...             & ...               & ...           \\
NGC~4696 & ...            & ...           & 3.5 $\pm$ 0.8    & 8.4 $\pm$ 0.9    & ...           & ...           & ...              & 6.4 $\pm$ 0.9   & ...               & 1.5 $\pm$ 0.1 \\
NGC~4697 & 14.5 $\pm$ 3.2 & 8.2 $\pm$ 1.1 & 7.4 $\pm$ 1.2    & 12.9 $\pm$ 1.1   & 0.3 $\pm$ 0.1 & 6.4 $\pm$ 0.5 & 8.2 $\pm$ 1.2    & 8.2 $\pm$ 2.0   & ...               & 1.6 $\pm$ 0.1 \\
NGC~5011 & ...            & ...           & ...              & 7.2 $\pm$ 1.0    & ...           & ...           & 2.1 $\pm$ 0.6    & 3.6 $\pm$ 0.6   & 1.8 $\pm$ 0.5     & ...           \\
NGC~5044 & 6.4 $\pm$ 1.3  & 1.8 $\pm$ 0.4 & 14.9 $\pm$ 0.9   & 14.6 $\pm$ 0.5   & ...           & 2.9 $\pm$ 0.2 & 1.8 $\pm$ 0.3    & 23.4 $\pm$ 1.1  & 1.2 $\pm$ 0.3     & 10.6 $\pm$ 0.4\\
NGC~5077 & ...            & ...           & 6.2 $\pm$ 1.1    & 9.1 $\pm$ 0.8    & 3.5 $\pm$ 0.6 & 1.7 $\pm$ 0.2 & 2.6 $\pm$ 0.6    & 9.4 $\pm$ 0.9   & 1.2 $\pm$ 0.5     & 2.8 $\pm$ 0.4 \\
NGC~5090 & ...            & 3.5 $\pm$ 0.9 & 7.0 $\pm$ 0.6    & 9.2 $\pm$ 0.8    & ...           & 1.8 $\pm$ 0.2 & 3.0 $\pm$ 0.8    & ...             & 3.1 $\pm$ 0.5     & ...           \\
NGC~5813 & ...            & 7.5 $\pm$ 1.5 & 6.4 $\pm$ 0.6    & 15.3 $\pm$ 1.3   & ...           & 2.9 $\pm$ 0.4 & 3.6 $\pm$ 1.0    & 6.3 $\pm$ 0.9   & ...               & ...          \\
NGC~5846 & ...            & ...           & ...              & ...              & ...           & ...           & ...              & 0.5 $\pm$ 0.1   & ...               & 0.6 $\pm$ 0.1 \\
NGC~5898 & ...            & ...           & ...              & ...              & ...           & ...           & ...              & 4.1 $\pm$ 1.4   & ...               & 0.9 $\pm$ 0.2 \\
NGC~6868 & 7.0 $\pm$ 2.0  & 3.5 $\pm$ 1.0 & 20.2 $\pm$ 1.4   & 16.3 $\pm$ 1.0   & ...           & 6.0 $\pm$ 0.5 & 4.0 $\pm$ 0.6    & 23.9 $\pm$ 1.5  & 1.4 $\pm$ 0.4     & 8.9 $\pm$ 0.3 \\
NGC~7079 & ...            & ...           & 2.1 $\pm$ 0.5    & 1.6 $\pm$ 0.3    & ...           & 1.0 $\pm$ 0.1 & 1.4 $\pm$ 0.3    & 5.0 $\pm$ 0.5   & 0.5 $\pm$ 0.1     & 1.2 $\pm$ 0.1 \\
NGC~7192 & ...            & ...           & ...              & 2.6 $\pm$ 0.8    & ...           & ...           & ...              & ...             & ...               & ...           \\
NGC~7332 & ...            & ...           & ...              & ...              & ...           & ...           & ...              & 3.9 $\pm$ 0.2   & ...               & ...           \\
IC~1459 & ...            & ...           & 5.0 $\pm$ 1.6    & 21.2 $\pm$ 1.5   & ...           & ...           & ...              & 10.3 $\pm$ 2.4  & ...               & 2.6 $\pm$ 0.5 \\
IC~3370 & ...            & ...           & 4.2 $\pm$ 0.3    & 4.8 $\pm$ 0.4    & ...           & ...           & 2.3 $\pm$ 0.3    & 3.7 $\pm$ 0.4   & ...               & 2.2 $\pm$ 0.2 \\
IC~4296 & ...            & ...           & ...              & ...              & ...           & ...           & ...              & ...             & ...               & 2.2 $\pm$ 0.5 \\
IC~5063 & ...            & ...           & 133.8 $\pm$ 27.7 & 106.5 $\pm$ 29.6 & ...           & ...           & 176.3 $\pm$ 44.8 & 244.9 $\pm$ 49.9& 788.5 $\pm$ 123.9 & ...         \\
\hline
\end{longtable}
\tablefoot{Values are in units of $10^{-18}$ W m$^{-2}$; uncertainties are 1 $\sigma$.
Neither atomic nor molecular emission lines are detected in Passive ETG.}
\end{landscape}
}

\onllongtabL{3}{
\begin{landscape}
\begin{longtable}{lccccccccccc}
\caption{Continued.}\label{tab3}\\
\hline
\hline
Galaxy  & $[$\ion{Ne}{II}] & $[$\ion{Ne}{V}] & $[$\ion{Ne}{III}] & H$_2$ 0--0 S(1) & $[$\ion{Fe}{II}] & $[$\ion{S}{III}] & $[$\ion{Ar}{III}] & $[$\ion{Ne}{V}] & $[$\ion{O}{IV}]\tablefootmark{a} & $[$\ion{Fe}{II}]\tablefootmark{a} \\
name & 12.81$\mu$m & 14.32$\mu$m & 15.55$\mu$m & 17.03$\mu$m & 17.94$\mu$m & 18.71$\mu$m & 21.83$\mu$m & 24.32$\mu$m & 
25.89$\mu$m & 25.99$\mu$m \\
\hline
NGC~1052 & 250.3 $\pm$ 9.7 & ...              & 180.4 $\pm$ 9.3   & 47.5 $\pm$ 9.7   & 43.6 $\pm$ 12.6 & 71.4 $\pm$ 9.4   & ...           & ...               & 27.2 $\pm$ 3.9   & 50.1 $\pm$ 9.3   \\
NGC~1209 & 3.1 $\pm$ 0.3   & ...              &  3.0 $\pm$ 0.2    & ...              &  0.9  $\pm$ 0.1 &  1.5 $\pm$ 0.1   & ...           & ...               & 0.4 $\pm$ 0.1    &  2.1 $\pm$ 0.2  \\
NGC~1297 & 1.9 $\pm$ 0.1   & ...              &  1.2 $\pm$ 0.1    & 7.5 $\pm$ 0.2    & ...             &  0.4 $\pm$ 0.1   & ...           & ...               & 0.8 $\pm$ 0.1    &  1.0 $\pm$ 0.1  \\
NGC~1366 & 1.9 $\pm$ 0.4   & ...              &  3.4 $\pm$ 0.4    & ...              & ...             & 1.7  $\pm$ 0.2   & ...           & ...               & 0.5 $\pm$ 0.1    &  0.5 $\pm$ 0.1  \\
NGC~1453 & 3.6 $\pm$ 0.4   & ...              & 7.5 $\pm$ 0.5     & 2.2 $\pm$ 0.2    & ...             & 2.3 $\pm$ 0.3    & ...           & ...               & ...              & ...             \\
NGC~1533 & 5.3 $\pm$ 0.3   & ...              & 6.6 $\pm$ 0.4     & 4.5 $\pm$ 0.2    & ...             & 5.7 $\pm$ 0.4    & 1.6 $\pm$ 0.2 & ...               & 0.6 $\pm$ 0.1    & 3.0 $\pm$ 0.3   \\
NGC~1553 &  25.1 $\pm$ 0.9  & 0.4 $\pm$ 0.1    & 22.3 $\pm$ 1.6    & 5.7 $\pm$ 0.5    & ...             & 8.5 $\pm$ 0.9    & ...           & ...               & 1.3 $\pm$ 0.2    & 6.1 $\pm$ 0.9   \\
NGC~2974 &  35.0 $\pm$ 1.0  & ...              & 27.5 $\pm$ 1.2    & 21.8 $\pm$ 1.1   & ...             & 13.3 $\pm$ 0.9   & ...           & 7.3 $\pm$ 0.9     & 2.1 $\pm$ 0.2    & 10.3 $\pm$ 0.6  \\
NGC~3258 & 15.2 $\pm$ 0.5  & ...              & 4.6 $\pm$ 0.3     & 6.6 $\pm$ 0.3    & 1.2 $\pm$ 0.2   & 3.7 $\pm$ 0.3    & 2.2 $\pm$ 0.5 & ...               & 1.1 $\pm$ 0.2    & 1.8 $\pm$ 0.4   \\
NGC~3268 & 10.5 $\pm$ 0.4  & ...              & 5.5 $\pm$ 0.2     & 3.0 $\pm$ 0.3    & 1.0 $\pm$ 0.3   & 3.4 $\pm$ 0.2    & 1.3 $\pm$ 0.2 & ...               & 1.0 $\pm$ 0.1    & 1.9 $\pm$ 0.2   \\
NGC~3557 & 9.6 $\pm$ 0.6   & ...              & 9.0 $\pm$ 0.6     & ...              & 1.4 $\pm$ 0.2   & 2.7 $\pm$ 0.3    & ...           & 1.1 $\pm$ 0.3     & 0.8 $\pm$ 0.1    & 2.8 $\pm$ 0.3   \\
NGC~3962 & 14.2 $\pm$ 0.4  & 1.0 $\pm$ 0.1    & 8.2 $\pm$ 0.5     & 4.7 $\pm$ 0.2    & 2.8 $\pm$ 0.2   & 5.3 $\pm$ 0.2    & ...           & ...               & 3.1 $\pm$ 0.2    & 3.0 $\pm$ 0.2   \\
NGC~4374 & 32.9 $\pm$ 1.3  & 1.6 $\pm$ 0.3    & 26.5 $\pm$ 1.1    & 8.3 $\pm$ 0.6    & 1.7 $\pm$ 0.3   & 14.9 $\pm$ 0.5   & 2.5 $\pm$ 0.6 & ...               & 3.5 $\pm$ 0.3    & 5.7 $\pm$ 0.6   \\
NGC~4552 & 13.5 $\pm$ 1.1  & 0.3 $\pm$ 0.1    & 10.4 $\pm$ 1.1    & ...              & ...             & 10.0 $\pm$ 1.0   & ...           & ...               & ...              & ...             \\
NGC~4636 &  12.9 $\pm$ 0.8  & ...              & 15.1 $\pm$ 0.8    & 2.4 $\pm$ 0.4    & ...             & 6.1 $\pm$ 0.5    & ...           & ...               & ...              & ...             \\
NGC~4696 & 13.3 $\pm$ 0.6  & ...              & 7.6 $\pm$ 0.3     & 6.4 $\pm$ 0.3    & ...             & 1.2 $\pm$ 0.1    & ...           & ...               & 1.5 $\pm$ 0.1    & 2.8 $\pm$ 0.2   \\
NGC~4697 & 8.3 $\pm$ 0.6   & ...              & 9.1 $\pm$ 1.1     & 4.0 $\pm$ 0.5    & ...             & 5.8 $\pm$ 0.8    & ...           & ...               & ...              & ...             \\
NGC~5011 & 2.8 $\pm$ 0.3   & 0.6 $\pm$ 0.2    & 3.9 $\pm$ 0.3     & ...              & ...             & 1.8 $\pm$ 0.1    & ...           & ...               & 0.9 $\pm$ 0.1    & 0.5 $\pm$ 0.1   \\
NGC~5044 & 24.3 $\pm$ 0.9  & ...              & 12.2 $\pm$ 0.4    & 13.4 $\pm$ 0.5   & 0.6 $\pm$ 0.1   & 3.1 $\pm$ 0.2    & 1.0 $\pm$ 0.2 & 1.2 $\pm$ 0.2     & 1.5 $\pm$ 0.1    & 2.1 $\pm$ 0.2   \\
NGC~5077 & 20.6 $\pm$ 1.0  & 0.6 $\pm$ 0.1    & 17.9 $\pm$ 0.8    & 12.7 $\pm$ 0.7   & 1.4 $\pm$ 0.2   & 6.3 $\pm$ 0.4    & 2.9 $\pm$ 0.6 & ...               & 1.7 $\pm$ 0.1    & 8.4 $\pm$ 0.5   \\
NGC~5090 &  20.7 $\pm$ 0.6  & 0.9 $\pm$ 0.1    & 9.4 $\pm$ 0.6     & 1.6 $\pm$ 0.3    & 3.8 $\pm$ 0.4   & 4.0 $\pm$ 0.3    & 4.1 $\pm$ 0.4 & ...               & 2.7 $\pm$ 0.3    & 2.2 $\pm$ 0.2   \\
NGC~5813 & 6.3 $\pm$ 0.3   & ...              & 7.0 $\pm$ 0.3     & 2.3 $\pm$ 0.4    & ...             & 3.0 $\pm$ 0.2    & ...           & ...               & ...              & ...             \\
NGC~5846 & 12.5 $\pm$ 0.9  & ...              & 9.1 $\pm$ 0.6     & 2.0 $\pm$ 0.2    & ...             & 3.8 $\pm$ 0.3    & ...           & ...               & ...              & ...             \\
NGC~5898 & 4.2 $\pm$ 0.3   & ...              & 8.3 $\pm$ 0.5     & 3.1 $\pm$ 0.3    & ...             & 3.8 $\pm$ 0.4    & ...           & 1.2 $\pm$ 0.4     & 0.5 $\pm$ 0.1    & 2.1 $\pm$ 0.2   \\
NGC~6868 & 30.0 $\pm$ 1.0  & 1.1 $\pm$ 0.1    & 23.8 $\pm$ 0.8    & 9.6 $\pm$ 0.6    & 1.5 $\pm$ 0.2   & 7.8 $\pm$ 0.4    & 1.3 $\pm$ 0.4 & 2.0 $\pm$ 0.4     & 4.8 $\pm$ 0.3    & 6.6 $\pm$ 0.4   \\
NGC~7079 & 2.9 $\pm$ 0.2   & ...              & 2.8 $\pm$ 0.2     & 3.7 $\pm$ 0.2    & ...             & 1.7 $\pm$ 0.1    & 2.3 $\pm$ 0.3 & 1.6 $\pm$ 0.2     & 1.3 $\pm$ 0.1    & 2.3 $\pm$ 0.2   \\
NGC~7192 & 2.4 $\pm$ 0.4   & ...              & 1.0 $\pm$ 0.2     & 0.3 $\pm$ 0.1    & 0.3 $\pm$ 0.1   & 0.3 $\pm$ 0.1    & 1.0 $\pm$ 0.2 & ...               & ...              & 0.6 $\pm$ 0.1   \\
NGC~7332 & ...             & ...              & 9.0 $\pm$ 0.9     & 1.7 $\pm$ 0.3    & ...             & 3.4 $\pm$ 0.5    & ...           & 1.2 $\pm$ 0.4     & 3.4 $\pm$ 0.2    & 4.7 $\pm$ 0.3   \\
IC~1459 & 51.4 $\pm$ 2.1  & ...              & 38.7 $\pm$ 1.7    & 1.4 $\pm$ 0.3    & 3.6 $\pm$ 0.6   & 9.9 $\pm$ 0.7    & ...           & 4.7 $\pm$ 1.1     & 2.8 $\pm$ 0.3    & 12.3 $\pm$ 1.0  \\
IC~3370 & 2.1 $\pm$ 0.1   & ...              & 1.4 $\pm$ 0.1     & 4.4 $\pm$ 0.2    & ...             & 0.6 $\pm$ 0.1    & ...           & 0.3 $\pm$ 0.1     & 0.7 $\pm$ 0.1    & 0.9 $\pm$ 0.1   \\
IC~4296 & 24.1 $\pm$ 1.4  & 0.5 $\pm$ 0.1    & 10.9 $\pm$ 0.7    & 1.7 $\pm$ 0.5    & 1.4 $\pm$ 0.3   & 4.2 $\pm$ 0.6    & ...           & ...               & 3.9 $\pm$ 0.4    & 2.9 $\pm$ 0.2   \\
IC~5063 & 238.7 $\pm$ 48.2& 335.2 $\pm$ 68.7 & 1103.6 $\pm$ 88.9 & 168.1 $\pm$ 36.9 & ...             & 321.7 $\pm$ 71.2 & ...           & 402.8 $\pm$ 107.1 & 477.1 $\pm$ 44.8 & 673.0 $\pm$ 51.4\\
\hline
\end{longtable}
\tablefoot{
\tablefoottext{a}{The two lines are blended in LL1 spectra: the values reported are the result
of a line de-blending.}}
\end{landscape}
}
\onllongtab{3}{
\begin{table*}
\caption{Continued.}
\centering
\begin{tabular}{lcccccccccccccc}
\hline
\hline
Galaxy &  H$_2$ 0--0 S(0) & $[$\ion{S}{III}] & $[$\ion{Si}{II}] & $[$\ion{Fe}{II}] & $[$\ion{Ne}{III}]\\
name &28.22$\mu$m& 33.48$\mu$m & 34.82$\mu$m & 35.35$\mu$m & 36.01$\mu$m \\
\hline
NGC~1052 & ...           & 84.6 $\pm$ 13.3  & 207.3 $\pm$ 18.4 & ...           & ...          \\
NGC~1209 & ...           &  3.5 $\pm$ 0.2   & 11.3 $\pm$ 0.3   & ...           & 1.1 $\pm$ 0.1\\
NGC~1297 & 3.0 $\pm$ 0.1 &  1.5 $\pm$ 0.1   &  5.5 $\pm$ 0.2   & 0.3 $\pm$ 0.1 & ...          \\
NGC~1366 & ...           & 1.5 $\pm$ 0.1    &  2.7 $\pm$ 0.2   & 1.5 $\pm$ 0.2 & ...          \\
NGC~1453 & ...           & ...              & ...              & ...           & ...          \\
NGC~1533 & 1.8 $\pm$ 0.3 & 6.3 $\pm$ 0.4    & 12.0 $\pm$ 0.4   & 0.5 $\pm$ 0.1 & ...          \\
NGC~1553 & ...           & 15.4 $\pm$ 1.0   & 28.9 $\pm$ 1.8   & ...           & 7.5 $\pm$ 0.7\\
NGC~2974 & 2.6 $\pm$ 0.6 & 13.5 $\pm$ 1.2   & 44.9 $\pm$ 1.2   & ...           & ...          \\
NGC~3258 & ...           & 5.6 $\pm$ 0.5    & ...              & ...           & ...          \\
NGC~3268 & ...           & 4.8 $\pm$ 0.3    & 7.6 $\pm$ 0.5    & ...           & ...          \\
NGC~3557 & ...           & 8.4 $\pm$ 0.4    & 7.1 $\pm$ 0.4    & 2.7 $\pm$ 0.3 & 4.3 $\pm$ 0.3\\
NGC~3962 & ...           & 6.2 $\pm$ 0.5    & 31.7 $\pm$ 1.0   & 1.4 $\pm$ 0.4 & 2.1 $\pm$ 0.3\\
NGC~4374 & 2.2 $\pm$ 0.6 & 24.3 $\pm$ 1.0   & 40.8 $\pm$ 1.3   & 3.6 $\pm$ 0.7 & ...          \\
NGC~4552 & ...           & ...              & ...              & ...           & ...          \\
NGC~4636 & ...           & ...              & ...              & ...           & ...          \\
NGC~4696 & 0.9 $\pm$ 0.1 & 4.9 $\pm$ 0.2    & 14.5 $\pm$ 0.5   & 4.0 $\pm$ 0.1 & 4.5 $\pm$ 0.2\\
NGC~4697 & ...           & ...              & ...              & ...           & ...          \\
NGC~5011 & ...           & 3.1 $\pm$ 0.2    & 5.0 $\pm$ 0.2    & ...           & ...          \\
NGC~5044 & 1.5 $\pm$ 0.2 & 9.0 $\pm$ 0.3    & 44.4 $\pm$ 1.1   & 0.8 $\pm$ 0.1 & 1.4 $\pm$ 0.1\\
NGC~5077 & 2.6 $\pm$ 0.5 & 11.7 $\pm$ 0.6   & 45.6 $\pm$ 1.4   & 2.1 $\pm$ 0.5 & 1.6 $\pm$ 0.5\\
NGC~5090 & 0.8 $\pm$ 0.2 & 5.3 $\pm$ 0.4    & 12.9 $\pm$ 0.5   & ...           & ...          \\
NGC~5813 & ...           & ...              & ...              & ...           & ...          \\
NGC~5846 & ...           & ...              & ...              & ...           & ...          \\
NGC~5898 & ...           & 4.5 $\pm$ 0.4    & 9.3 $\pm$ 0.5    & ...           & 2.0 $\pm$ 0.4\\
NGC~6868 & ...           & 22.3 $\pm$ 0.8   & 28.8 $\pm$ 1.0   & 2.8 $\pm$ 0.5 & 3.8 $\pm$ 0.5\\
NGC~7079 & 1.0 $\pm$ 0.2 & 5.2 $\pm$ 0.4    & 5.9 $\pm$ 0.4    & 2.1 $\pm$ 0.3 & 1.7 $\pm$ 0.4\\
NGC~7192 & ...           & 2.7 $\pm$ 0.2    & 4.0 $\pm$ 0.2    & ...           & ...          \\
NGC~7332 & 1.8 $\pm$ 0.3 & 4.7 $\pm$ 0.3    & 8.2 $\pm$ 0.7    & 1.0 $\pm$ 0.3 & 1.1 $\pm$ 0.2\\
IC~1459 & 3.1 $\pm$ 1.0 & 23.5 $\pm$ 1.7   & 50.0 $\pm$ 2.2   & ...           & ...          \\
IC~3370 & 0.7 $\pm$ 0.1 & 1.6 $\pm$ 0.2    & 0.9 $\pm$ 0.2    & 1.5 $\pm$ 0.2 & ...          \\
IC~4296 & ...           & 10.1 $\pm$ 1.1   & 15.0 $\pm$ 0.8   & 2.0 $\pm$ 0.5 & ...          \\
IC~5063 & ...           & 480.5 $\pm$ 58.9 & 524.9 $\pm$ 66.5 & ...           & ...          \\
\hline
\end{tabular}
\end{table*}
}


\begin{table*}
\caption{PAH complex intensities.}
\centering
\begin{tabular}{lcccccc}
\hline
\hline
Galaxy   &  6.22$\mu$m  &   7.7$\mu$m\tablefootmark{a} &     8.6$\mu$m &      11.3$\mu$m\tablefootmark{b} &
  12.7$\mu$m\tablefootmark{c}  &   17$\mu$m\tablefootmark{d} \\
name &{($10^{-18}$ W m$^{-2}$)}&{($10^{-18}$ W m$^{-2}$)}&{($10^{-18}$ W m$^{-2}$)}&{($10^{-18}$ W m$^{-2}$)}&
{($10^{-18}$ W m$^{-2}$)}&{($10^{-18}$ W m$^{-2}$)}\\
\hline
NGC~1052 & 120.0 $\pm$ 33.0 & 239.0 $\pm$ 41.4 & ...            & 195.8 $\pm$ 40.4 & ...             & 254.1 $\pm$ 66.3 \\
NGC~1297 & 13.3 $\pm$ 4.4   & 62.9 $\pm$ 3.3   & 13.8 $\pm$ 1.0 & 44.9 $\pm$ 1.4   & 21.8 $\pm$ 1.3  & 23.6 $\pm$ 2.5   \\
NGC~1453 & ...              & 77.4 $\pm$ 10.7  & 10.6 $\pm$ 2.4 & 31.0 $\pm$ 2.8   & 20.8 $\pm$2.1   & 13.2 $\pm$ 1.7\\
NGC~1533 & ...              & ...              & ...            & 31.3 $\pm$ 3.8   & 12.6 $\pm$2.1   & 27.4 $\pm$ 1.9\\
NGC~1553 & 105.6 $\pm$ 10.3 & 406.8 $\pm$ 35.3 & 48.7 $\pm$ 7.4 & 275.8 $\pm$ 10.5 & 84.4 $\pm$7.0   & 78.3 $\pm$ 5.5\\
NGC~2974 & 88.3 $\pm$ 9.4   & 409.7 $\pm$ 21.2 & 44.0 $\pm$ 3.9 & 256.9 $\pm$ 7.2  & 113.5 $\pm$ 5.7 &144.9 $\pm$ 5.3\\
NGC~3258 & 113.8 $\pm$ 4.9  & 537.6 $\pm$ 15.3 &110.4 $\pm$ 4.6 & 119.4 $\pm$ 3.1  & 72.0 $\pm$ 2.9  & 44.1 $\pm$ 1.4\\
NGC~3268 & 67.5 $\pm$ 6.4   & 341.1 $\pm$ 9.6  & 48.3 $\pm$ 2.9 & 150.5 $\pm$ 4.3  & 83.6 $\pm$ 4.1  &73.3 $\pm$ 2.5\\
NGC~3557 & 32.5 $\pm$ 10.7  & 134.9 $\pm$ 11.5 & 19.6 $\pm$ 5.8 & 91.5 $\pm$ 4.9   & 48.4 $\pm$ 5.8  &32.5 $\pm$ 3.9\\
NGC~3962 & 31.9 $\pm$ 8.8   & 133.1 $\pm$ 8.8  & 24.6 $\pm$ 3.8 & 102.8 $\pm$ 4.6  & 37.3 $\pm$ 2.9  & 35.3 $\pm$ 4.8\\
NGC~4374 & 65.7 $\pm$ 20.6  & 258.3 $\pm$ 28.7 & 51.3 $\pm$ 6.6 & 178.9 $\pm$ 9.4  & 95.1 $\pm$ 7.0  & 65.2 $\pm$ 4.7\\
NGC~4552 & ...              & ...              & ...            & 38.3 $\pm$ 7.9   & ...             & 7.9 $\pm$ 1.8 \\
NGC~4636 & ...              & ...              & ...            & 20.7 $\pm$ 5.0   & ...             & 8.8 $\pm$ 0.8\\
NGC~4696 & ...              & ...              & ...            & 11.3 $\pm$ 1.7   & ...             & 7.5 $\pm$ 0.5\\
NGC~4697 & 211.0 $\pm$ 21.2 & 816.8 $\pm$ 44.5 &199.0 $\pm$ 16.2&263.0 $\pm$ 14.2  & 179.3 $\pm$ 5.0 &78.4 $\pm$ 7.9\\
NGC~5044 & 29.4 $\pm$ 7.0   & 112.5 $\pm$ 5.8  & 32.1$\pm$3.1   & 58.1 $\pm$ 2.1   & 27.1 $\pm$ 1.8  &28.8 $\pm$ 0.9\\
NGC~5077 & ...              & 66.6 $\pm$ 6.5   & 22.3$\pm$4.7   & 30.8 $\pm$ 2.5   & 14.0 $\pm$ 1.8  &41.2 $\pm$ 2.6\\
NGC~5090 & 13.5 $\pm$ 4.1   & 80.6 $\pm$ 13.4  & 10.5$\pm$1.6   & 34.5 $\pm$ 2.4   & 21.8 $\pm$ 2.4  & 11.7 $\pm$ 3.1\\
NGC~5898 & ...              & 53.0 $\pm$ 16.3  & ...            & 46.5 $\pm$ 4.3   & 15.1 $\pm$ 3.3  & 21.7 $\pm$ 2.4\\
NGC~6868 & 50.5 $\pm$ 12.3  & 242.7 $\pm$ 24.3 & 27.5 $\pm$ 2.9 & 129.6 $\pm$ 6.9  &  86.8 $\pm$ 4.9 & 45.6 $\pm$ 3.1\\
NGC~7079 & 23.1 $\pm$ 4.9   & 90.2 $\pm$ 3.9   & 24.7 $\pm$ 3.0 & 65.5 $\pm$ 2.5   &29.8 $\pm$ 2.3   &20.4 $\pm$ 0.6\\
IC~1459  & ...              & 108.4 $\pm$ 15.1 & 6.1 $\pm$ 1.8  & 84.9 $\pm$ 6.0   & 41.0 $\pm$ 5.4  &47.1 $\pm$ 4.7\\
IC~3370  & ...              & 62.9 $\pm$ 19.6  & 13.1 $\pm$ 3.5 & 22.2 $\pm$ 3.6   &  22.5 $\pm$ 5.7 &21.5 $\pm$ 2.5\\
IC~5063  & ...              &1190.4 $\pm$ 140.9&208.8 $\pm$ 68.9&462.9 $\pm$ 125.4 & ...             & ... \\
IC~4296  & ...              & ...              & ...            &  26.7$\pm$5.8    &  26.0 $\pm$ 5.6 & 19.8 $\pm$ 5.6\\
\hline
\end{tabular}
\label{tab4}
\tablefoot{uncertainties are 1 $\sigma$.
\tablefoottext{a}{Includes the 7.42~$\mu$m, 7.60~$\mu$m and the 7.85~$\mu$m features.}
\tablefoottext{b}{Includes the 11.23~$\mu$m, and the 11.33~$\mu$m features.} 
\tablefoottext{c}{Includes the 12.62~$\mu$m and the 12.69~$\mu$m features.}
\tablefoottext{d}{Includes the 16.45~$\mu$m, 17.04~$\mu$m, 17.375~$\mu$m and the 17.87~$\mu$m features.}
}
\end{table*}



\section{Observations and data reduction}
\label{sec:data}

The data set presented here is composed of MIR spectra obtained with the 
Infrared Spectrograph instrument of the {\it Spitzer} space telescope. It is composed of 
data from our own proposal (PI Rampazzo in Table 2, program ID 30256), 
obtained during the third {\it Spitzer} General Observer Cycle on 2007 June 1, 
and data retrieved from the {\it Spitzer} archive 
(mainly from Bregman, ID 3535, and Kaneda, ID 3619 and ID 30483).
The details of the observations for each galaxy are provided in Table~\ref{tab2}.

Our observations were performed in Standard Staring mode with low
resolution ($R\sim$ 64--128) modules SL1 (7.4--14.5$\mu$m), SL2
(5--8.7$\mu$m), LL2 (14.1--21.3$\mu$m) and LL1 (19.5--38$\mu$m).
Observations of the other investigators do not include, in general, 
all IRS modules.

The data reduction used for these data was first presented in \citet{Bressan06},
but was optimized later, as described below.

We first removed bad pixels from the IRS coadded images with an adapted version of the
{\tt IRSCLEAN}\footnote{http://ssc.spitzer.caltech.edu/archanaly/contributed/irsclean/}
procedure. Then, the sky background was removed by subtracting coadded
images taken with the source placed in different orders at the same nod position.
For most extended galaxies (IC~1459, NGC~1052, NGC~1407, NGC~1553, NGC~2974, NGC~3557, NGC~3962, 
NGC~4696, NGC~5090) offset exposures were obtained to measure the sky background
without contamination from the target galaxy itself. For those galaxies for which only
LL2 module was used, the background subtraction was done by subtracting coadded
images taken with the source at different nod positions.

In order to derive calibrated spectral energy distributions, we have to take into account
that the galaxies in the sample are extended, compared to the IRS point spread function (PSF).
Since the IRS spectra are calibrated on point-sources,
we have devised an {\it ad hoc} procedure to correct for the effects of 
the variation with the wavelength of the IRS PSF. This procedure
exploits the large degree of symmetry
characterizing the light distribution in ETGs.
This procedure was applied to SL modules but not to LL modules where the
sources can be considered as point sources with respect to the IRS PSF.

We obtained new e$^-$~s$^{-1}$ to Jy flux conversions by applying a correction
for aperture losses (ALCF), and a correction for slit losses (SLCF), to
the flux conversion tables provided by the Spitzer Science Center
\citep{Kennicutt03}. By applying the ALCF and SLCF corrections, we
obtained the flux received by the slit.

For each galaxy, we simulated the corresponding observed
linear profile along the slits by convolving with a wavelength dependent
two-dimensional intrinsic surface brightness profile with the instrumental
PSF. The adopted profile is a two dimensional
modified King law \citep{Elson87}. By fitting the observed profiles with
the simulated ones, we can reconstruct the intrinsic profiles and the
corresponding intrinsic spectral energy distribution (SED).
This procedure also has the advantage of determining whether a particular 
feature is spatially extended or not.

Finally the spectrum was extracted in a fixed width of 18\arcsec\ for SL
around the maximum intensity (corresponding to an aperture
of 3\farcs 6 $\times$ 18\arcsec). The LL spectrum was
rescaled to match that of SL.

The uncertainty in the flux was evaluated by considering the dark current
noise, the readout noise, and the Poissonian noise from the background and from
the source. The Poissonian noise associated with the sources' flux was
estimated as the square root of the ratio between the variance of the 
number of e$^-$ extracted per pixel in each exposure, and the number
of the exposures. The same method was applied to derive the Poissonian noise
from the background; where background was extracted from the coadded images
in the IRS order where the source was not present.

We notice that the overall absolute photometric uncertainty of IRS is 10\%,
while the slope deviation within a single segment (affecting all spectra
in the same way) is less than 3\% (see the Spitzer Observer Manual).

The rest frame flux calibrated IRS spectrum of each galaxy is shown in 
Fig.~\ref{Fig1}. The MIR spectra of our ETGs
encompass a wide range of morphologies: AGN like,
with PAH emission characteristic of star-forming galaxies,
with unusual PAH emission, with only  ionic/molecular emission lines,
and finally, typical of passively evolving ETGs.

The following sections are devoted to the quantitative analysis of these
spectra.


\section{Analysis and classification of MIR spectra} 
\label{sec:analysis}

Although the ETGs of this sample have quite similar
optical spectra (with most of them classified from 
optical emission lines as LINERs), the
MIR spectra show a surprising diversity.

\begin{figure}
\centering
\includegraphics[width=0.45\textwidth]{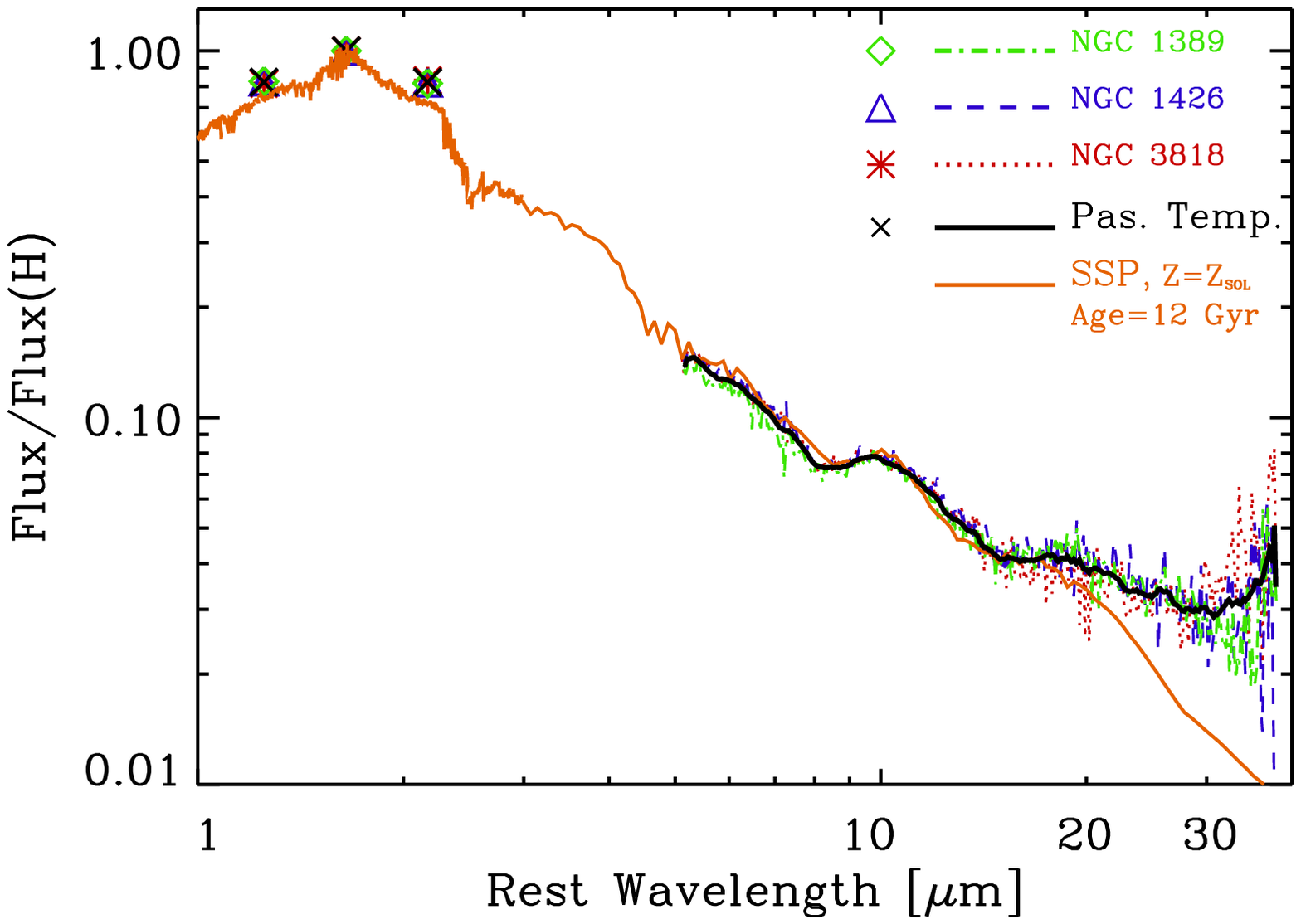}
\caption{The passive template adopted in this paper (black solid line, and black crosses 
in the plot), 
is compared with the \textit{IRS-Spitzer} spectra (lines) of the three galaxies used to
build it, and with an SSP of solar metallicity and age of 12 Gyr. 
The spectra, at the rest wavelength, are normalized to the H-band.  The 2MASS 
H, J, K-bands fluxes of the three galaxies, within the central 5\arcsec\ radius, are indicated
with symbols in the top left of the plot. At wavelengths longer than $\approx$30 $\mu$m 
there is a residual contribution of diffuse warm dust which does not affect our analysis 
(see text).}
\label{Fig2}%
\end{figure}

\subsection{Passive ETGs}

We can, as a first step, separate the spectra into "passive" and "active" \citep[see][]{Bressan06}, where
"passive" spectra are those that do not show any emission lines or PAH 
emission features, while "active" refers to objects that show one or the other, or both.

In our classification scheme we define as \textit{Class--0} those ETGs with purely passive spectra.
Neither ionic or molecular emission lines, nor PAHs, are revealed in the
spectra of NGC~1389, NGC~1407, NGC~1426, NGC~3818, NGC~5638, 
NGC~5812, NGC~5831, and IC~2006 \citep[see also][]{Kaneda08,Bregman06}.
The spectra of IC~2006, NGC~1407, NGC 5638, NGC 5812, and NGC 5831 
have, however, a low signal-to-noise ratio. Class~0 accounts for $\approx$20\% of the  
total MIR sample. Notice from Table~5 that, with the exclusion of IC~2006,
{\it Class--0} ETGs show no evidence of emission lines in their optical spectra, 
and were classified as {\it Inactive}. 
In the optical, IC~2006 exhibits weak emission lines, and is classified as a 
composite/transition object.
The MIR spectrum of this ETG indicates a possible detection of the [\ion{Ne}{III}]15.55$\mu$m 
emission line  at $< 2 \sigma$ level (see Fig.~\ref{Fig4}). However, due to the 
low signal-to-noise ratio of the spectrum, we classify the MIR spectrum of this galaxy as purely passive.

We averaged the spectra of the three best representatives of Class-0 objects
(namely NGC~1389, NGC~1426, and NGC~3818), 
obtaining a template spectrum representative of an old stellar population.
The MIR galaxy spectra were further extended into the NIR spectral region
with 2MASS fluxes within the central 5\arcsec\ radius (similar to the IRS 
extraction aperture), and then normalized in the H-band.
The template and the spectra of the three ETGs are shown in Fig.~\ref{Fig2}.
In the same figure we show also the SED of an SSP of solar metallicity and
age 12 Gyr \citep[from][]{Bressan98}.

The main characteristic of passive spectra is the presence of the broad emission feature 
around $\sim$10~$\mu$m that is attributed to silicate emission arising from the
dusty circumstellar envelopes of O-rich AGB stars 
\citep[see][]{Athey02,Molster02,Sloan98,Bressan98} superimposed on the photospheric stellar continuum
from red giant stars. The spectra also clearly show the presence of a less pronounced bump at around
18 $\mu$m, likely arising from the same silicate circumstellar dust as predicted by
\citet{Bressan98} and as shown in the SSP SED. 
An hint of the presence of this bump can be found in the spectra presented in
\citet{Bressan06}, although it is only clear in the present spectra that cover a larger
spectral window. Finally we note the presence of a dip at 8 $\mu$m probably due to
photospheric SiO absorption bands \citep[e.g.][]{Verhoelst09}. 

It is worth noticing that in this process we are assuming that the old 
stellar population is similar in all the galaxies analyzed. 
Indeed, inspection of Fig. \ref{Fig2} reveals that, once normalized to the
2MASS flux, the IRS spectra of the three passive galaxies are very similar, especially in the region
where the old population dominates, below $\sim$15~$\mu$m. At wavelengths 
longer than $\sim$30~$\mu$m the spectra show a rising continuum which
could be due to the presence of diffuse dust even in these  ``passive'' objects.
However, this will not affect our analysis because this rise is negligible with
respect to other contributions that dominate this spectral region in ``active'' ETGs.


\subsection{Measurement of the emission lines and PAH features}
\label{sec:emlines}

Intensities of emission lines and  PAH features were obtained from the extracted spectra with an
ad-hoc algorithm devised to decompose the \textit{Spitzer-IRS} spectra of ETGs \citep{Vega10}.
The algorithm is similar to the {\tt PAHFIT} tool \citep{Smith07}, 
but  with the noticeable difference that it includes a more adequate treatment 
of the emission from the old stellar population, whose contribution to the 
MIR spectra of ETGs cannot be neglected.
This contribution can be of the order of  99\% at the shorter wavelengths,    
$\lambda \sim 5.6 \mu$m, and may remain important also at longer wavelengths 
\citep{Vega10}.

\begin{figure}
\centering
\includegraphics[width=0.45\textwidth]{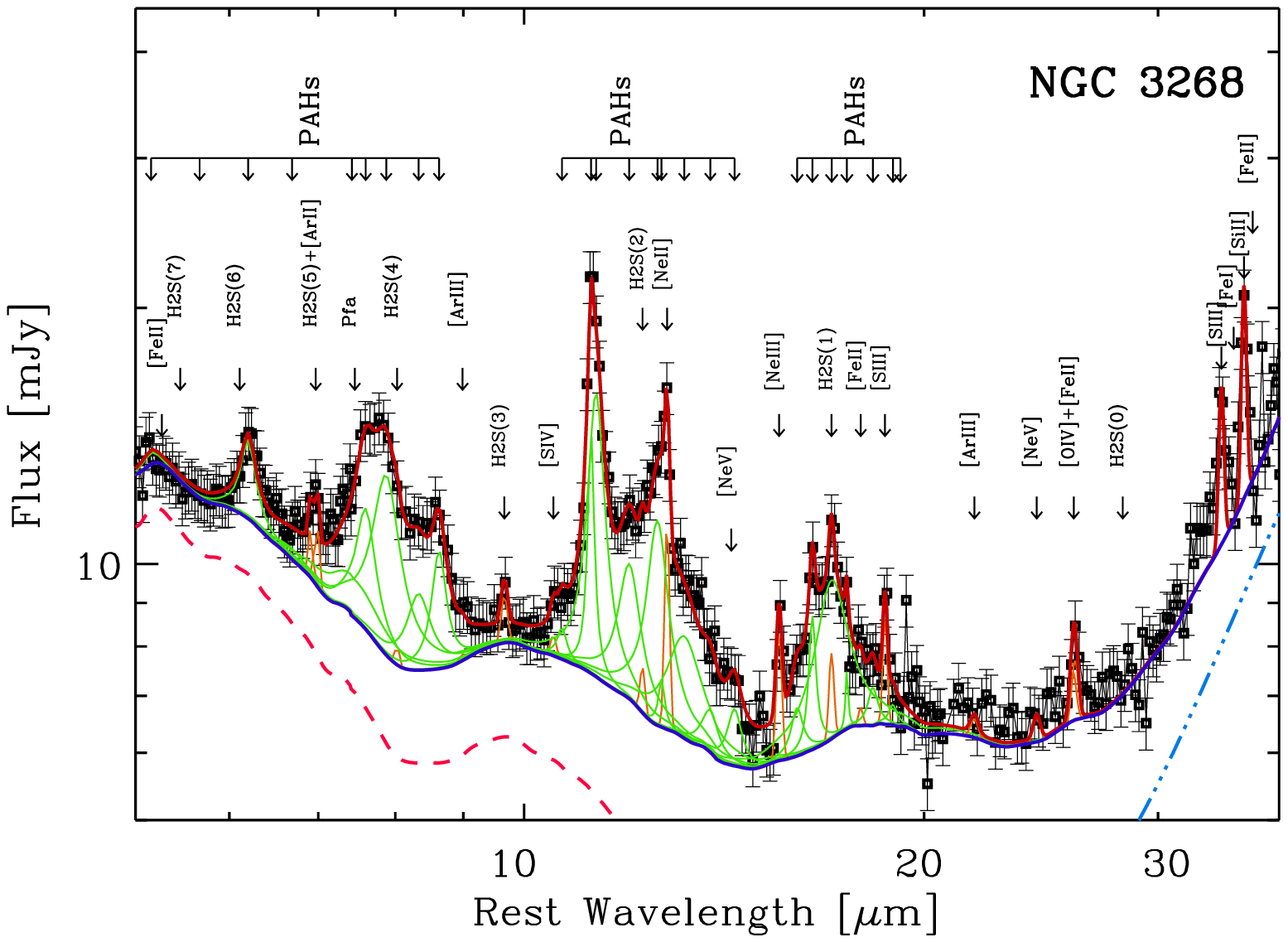}
\includegraphics[width=0.45\textwidth]{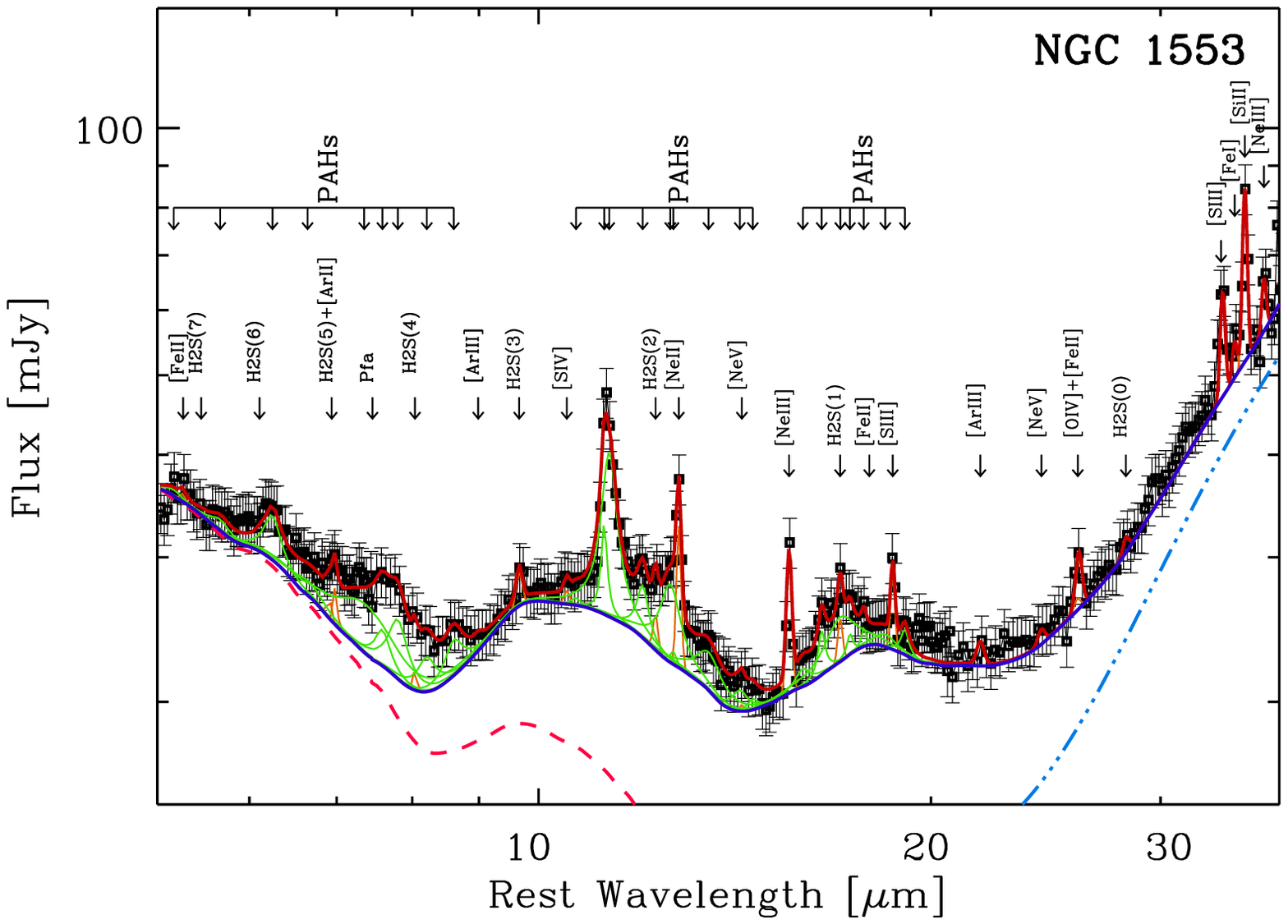}
\includegraphics[width=0.45\textwidth]{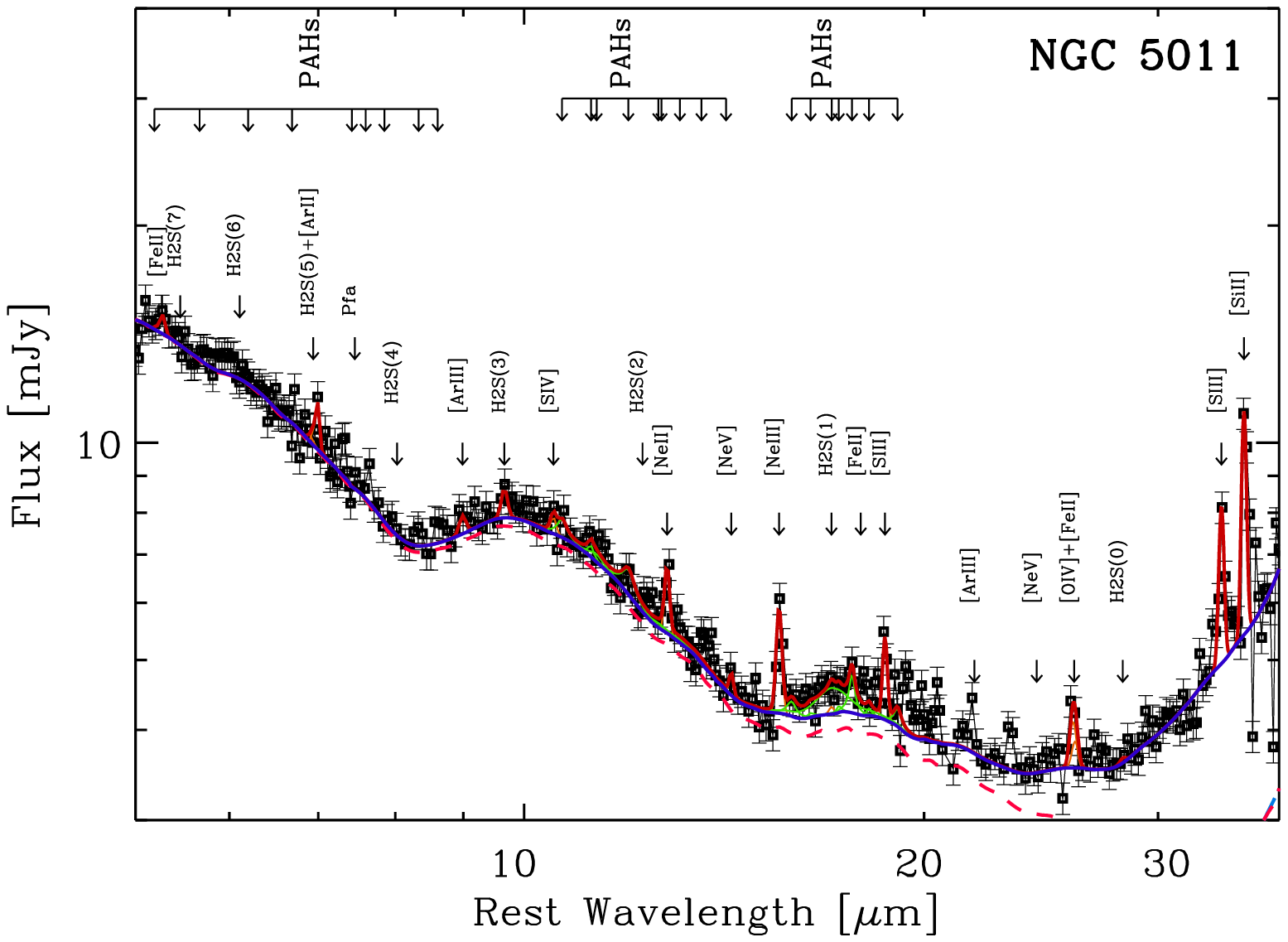}
\caption{{\em Upper panel:} Detailed fit of the MIR spectrum of a typical ETG with normal PAH 
emission. Open squares and the solid thick red line are the observed MIR 
spectra and our final best fit, respectively. The fit is calculated
as the sum of an underlying continuum (solid thick blue line),
the PAH features (solid thin green line) and the emission lines (solid thin orange line).
The two components of the continuum, old stellar population (dashed line)
and diffuse dust emission (dot-dashed line), are also plotted.
{\em Middle panel:} As before but for an ETG with unusual PAH emission.
{\em Bottom panel:} As before but for an ETG showing only line emission.
}
\label{Fig3}%
\end{figure}

    
The contribution due to the old  stellar population has been removed 
from the other galaxies by subtracting the template after normalization 
to the corresponding 2MASS H-band flux, in the same  5\arcsec\ radius aperture.
In this way we find that, for the bulk of the sample (35/40), 
the contribution of the old stellar component to the observed fluxes at 
5.6~$\mu$m is greater than 95\%. The contribution is lower in
NGC~3258 ($\sim$88\%), NGC~4697 ($\sim$87\%) and NGC~3256 ($\sim$85\%).

The subtracted (residual) spectra were fitted by a combination of modified 
black bodies describing dust emission, Drude profiles for PAH features, and 
Gaussian profiles for atomic and H$_2$ emission  lines, as fully detailed in 
\citet{Vega10}.
In NGC~1052 and IC~5063, where the contribution from the old population is almost 
negligible ($\sim$19\% and $\sim$3\%, respectively), and the spectrum is typical of an 
AGN \citep{Wu09}, we used suitable spline functions to represent the underlying continuum.

Figure~\ref{Fig3} show examples of the spectral decomposition
in three typical galaxies with different degrees of activity.

The intensities of the MIR emission lines for the whole sample
are listed in Table~\ref{tab3} (this table is available electronically only). 
Table \ref{tab4} lists the intensities of the main PAH-complexes, for all 
the galaxies where they were detected. The decomposition of PAH complexes 
into individual PAH features is provided only in electronic form.

\begin{table}
\caption {Summary of the optical and  {\it Spitzer}--IRS analyses}
\centering
\begin{tabular}{lcc}
\hline
\hline
Ident. &Activity   &Spectrum\\
       &class  Opt. &Morph. MIR  \\
\hline
NGC~1052 &LIN(H) &4 \\
NGC~1209 &LIN(H)&1\\
NGC~1297 &LIN(H)&2 \\
NGC~1366 &IN&1\\
NGC~1389 &IN&0\\
NGC~1407 &IN&0\\
NGC~1426 &IN&0\\
NGC~1453 &LIN(H)&2\\
NGC~1533 &LIN(H)&2 \\
NGC~1553 &LIN(W)&2 \\
NGC~2974 &LIN(H)&2 \\
NGC~3258 &Comp(H)&3 \\
NGC~3268 &LIN(H)&3 \\
NGC~3557 &LIN(W)&2 \\
NGC~3818 &IN(Traces)&0 \\
NGC~3962 &LIN(H)&2 \\
NGC~4374 &LIN(H)&2 \\
NGC~4552 &Comp(W)&2 \\
NGC~4636 &LIN(H)&2 \\
NGC~4696 &LIN(H)&2 \\
NGC~4697 &LIN(W)&3 \\
NGC~5011 &LIN(W)&1 \\
NGC~5044 &LIN(H)&2 \\
NGC~5077 &LIN(H)&2 \\
NGC~5090 &LIN(H)&2 \\
NGC~5638 &IN&0 \\
NGC~5812 &IN&0 \\
NGC~5813 &LIN(W)&1 \\
NGC~5831 &IN&0 \\
NGC~5846 &LIN(H)&1 \\
NGC~5898 &LIN(W)&2 \\
NGC~6868 &LIN(H)&2 \\
NGC~7079 &LIN(W)&2 \\
NGC~7192 &LIN(W)&1 \\
NGC~7332 &IN(Traces)&1 \\
IC~1459 &LIN(H)&2 \\
IC~2006 &Comp(W)&0 \\
IC~3370 &LIN(H)&2 \\
IC~4296 &LIN(H)&2 \\
IC~5063 &AGN&4 \\
\hline
\end{tabular}
\label{tab5}
\tablefoot{
Summary of the present observations and comparison with \citetalias{Annibali10a}.
The optical activity class (column 2: \citetalias{Annibali10a}) uses the following
notation: LIN = LINER; AGN = AGN like emission; IN = either faint 
(Traces) or no emission lines; Comp = transition between HII regions and LINERs. 
W and H  indicate weak emission ($EW$(H$\alpha +$[\ion{N}{II}]$\lambda$6584)$<3 \AA$) and 
strong emission line galaxies, respectively.
In column 3 we report the MIR spectrum morphology discussed in Sect.~\ref{sec:analysis}.}
\end{table}

\subsection{ETGs with emission lines but without PAH features}

Showing only a possible faint [\ion{Ne}{III}]15.55$\mu$m atomic emission line, 
IC~2006 (belonging to Class--0) could mark a sort of transition  between {\it purely passive}  
and {\it ETGs with emission lines but without PAH features}, or  \textit{Class--1} 
ETGs, in our scheme (see Fig.~\ref{Fig5}).
The continuum of all these ETGs is the same of the passive template up to
25~$\mu$m, showing in some cases more dust emission at longer wavelenghts.
Ionic (in particular [\ion{Ne}{II}]12.8$\mu$m,
[\ion{Ne}{III}]15.55$\mu$m, [\ion{S}{III}]12.82$\mu$m lines) and
H$_2$ rotational lines (in particular S(1)$\lambda$17.0) are detected. 
Still, no PAHs are revealed in NGC~1209, NGC~1366, NGC~5011,
NGC~5813,  NGC~5846, NGC~7192, or NGC~7332.  
This class includes $\approx$22\% of our ETGs. Most of Class--1
ETGs in the MIR are LINERs in the optical region. The only exception is
NGC~1366 which has been classified {\it Inactive} in the optical.

None of the ETGs in {\it Class--0} and {\it Class--1} can be arranged into 
the A, B, C, D classes of LINERs devised by \citet{Sturm06},
since all of these latter show PAH complexes. 

\subsection{ETGs with PAHs}

PAH  emission has been detected in many ETGs
\citep[see e.g.][]{Bregman06,Bressan06,Kaneda05,Panuzzo07,Kaneda08}.  
We detect PAHs in $\approx$62.5\% of our MIR sample.

ETGs showing PAHs are collected in Figs.~6~--~9. The MIR spectra also exhibit  
forbidden nebular emission lines of several elements
like Ar, Fe, Ne, O, S, and Si (see Table~\ref{tab3}). The galaxies 
displayed in Fig.~\ref{Fig6} show a prominent 11.3 $\mu$m PAH feature 
and  have {\it unusual 7.7$\mu$m/11.3$\mu$m PAH band ratios}, 
typically $\lesssim 2.3$. 
They represent 50\% of our total sample, and  80\% of 
the sub-set of ETGs with PAH features.  We classify these galaxies, 
with unusual PAH ratios, as \textit{Class--2} ETGs.
Within this class we include also NGC~4636 and  NGC~4696,  
which show very faint 6.3$\mu$m, 7.7$\mu$m and 8.6$\mu$m
PAH emission (see Table~4), although 11.3$\mu$ m PAH in both cases is detected.
Notice that for NGC~4696 \citet{Kaneda08} report  inter-band strength ratios 
of 7.7$\mu$m/11.3$\mu$m $<$0.6 and 17$\mu$m/11.3$\mu$m $< 0.88\pm$0.5.

Figure~\ref{Fig7} shows  {\it Class--2} ETGs whose spectra, 
in addition to PAH features, exhibit prominent  H$_2$ rotational emission
lines. In NGC~1297, NGC~5044, NGC~5077, NGC~6868, and NGC~7079  most
or even the full series of emission lines from H$_2~0-0$ S(0)-S(7) are detected.

NGC~3258,  NGC~3268, and NGC~4697 are shown separately 
in Fig.~\ref{Fig8}, since their spectra are dominated by 7.6$\mu$m, 11.3$\mu$m,
12.7$\mu$m and 17$\mu$m PAHs, and {\it present normal PAH inter-band 
ratios} \citep[i.e. $> 2.5$, ][]{Lu03,Smith07}. 
Their spectra are very similar to the post-starburst
ETG NGC~4435 \citep{Panuzzo07}, 
shown in the same plot for comparison. The shape of the spectra is reminiscent 
of the 1B class in \citet{Spoon07}, i.e. a MIR spectrum exhibiting the 
family of PAH features at 6.2, 7.7, 8.6, 11.2, 12.7, and 17.3$\mu$m 
on top of a hot dust continuum. Prototypical of the Class~1~B in 
Spoon et al. is the nucleus of the Seyfert 1.2, barred spiral NGC~7714. 
However, NGC~3258 and NGC~3268 also present  a steepening 
of the 20-30 $\mu$m continuum, which is absent in the 1 B class but 
present in the 2C class, and which could be due to a cold dust component.
We place these galaxies with ``normal'' PAH emission ratios into \textit{Class--3}.

\subsection{ETGs with hot dust continuum}

The MIR spectra of NGC~1052, considered a prototypical LINER 
\citep[see e.g.][]{Ho08}, and of IC~5063, a well known Seyfert 
(see for references on-line notes in R05) are shown in Fig.~\ref{Fig9}.
The continuum of both galaxies is dominated by hot dust, most probably
coming from an AGN torus.
PAH emission features are also visible; in particular, NGC~1052
shows both the 7.6 and 11.3$\mu$m PAH complexes, while 
it is possible to distinguish only the 7.6$\mu$m feature in IC~5063.
H$_2$ 0-0 S(1) and S(3) molecular emission lines are present in both
spectra. Ionic forbidden emission lines of [\ion{Ar}{II}], [\ion{Ne}{II}] and
[\ion{Ne}{III}], [\ion{S}{III}] and [\ion{O}{IV}] are revealed in both spectra.
High-ionization [\ion{S}{IV}] and [\ion{Ne}{V}] emission lines are
revealed only in IC~5063. We classify the MIR spectra of these two
ETGs as \textit{Class--4}.

In column 3 of Table~5 we report the MIR spectrum morphology.
In Table 6 we summarize the fraction of our ETGs in the different
classes.


\begin{table}
\caption{ETGs in the different classes of our classification
scheme.}
\centering
\begin{tabular}{lccccc}
\hline
\hline
   & Class--0  & Class--1 & Class--2 & Class--3 & Class--4  \\
\hline
number &   8  & 7   & 20  &  3  & 2 \\
\%          & 20 & 17.5 &50 & 7.5 &5 \\
\hline
\end{tabular}
\label{tab6}
\end{table}


\begin{figure*}
 \centering
 \includegraphics[width=12.5cm]{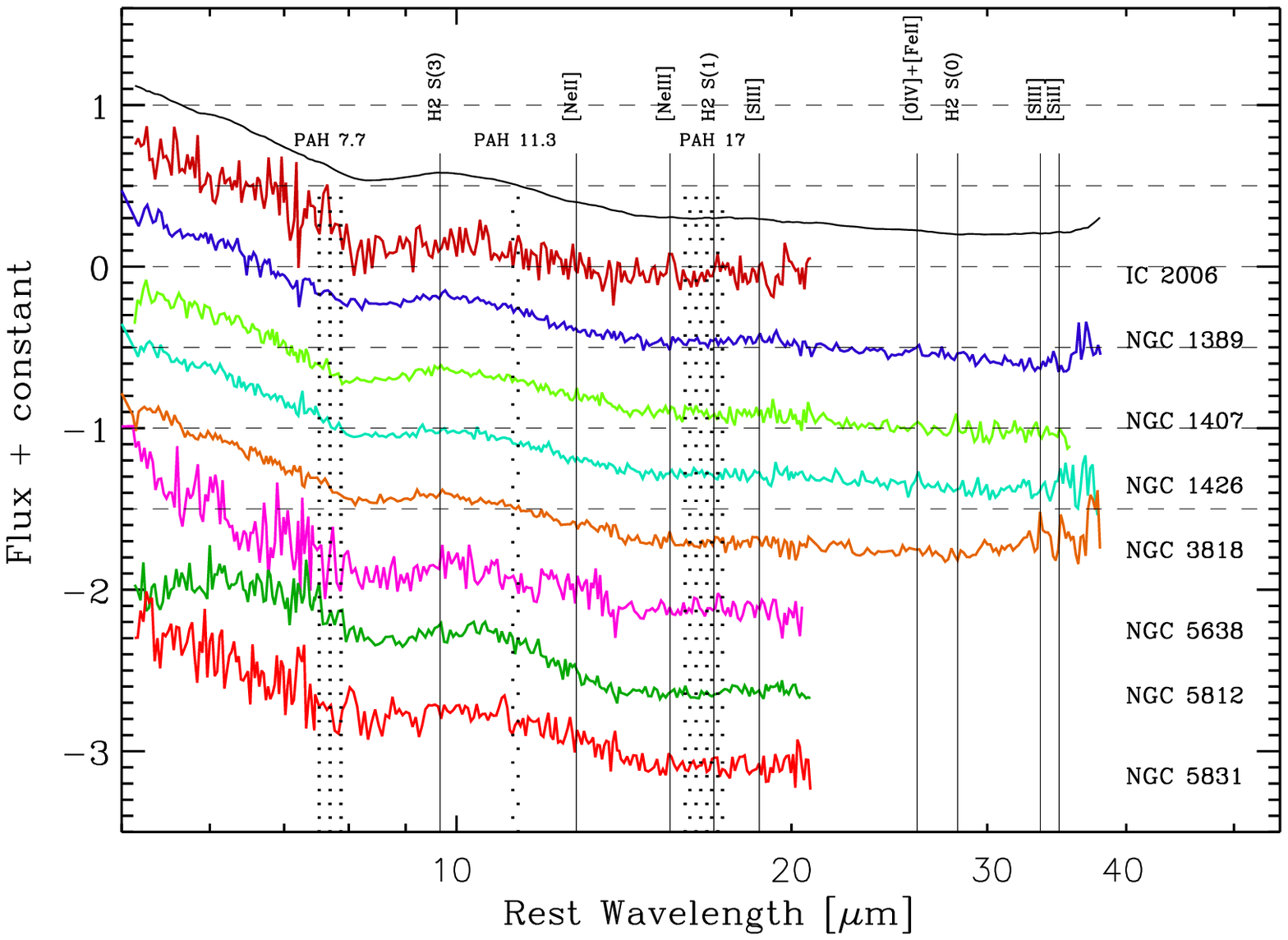}
   \caption{{\it Class--0} ETGs i.e. showing neither PAH nor ionic or molecular
   emission lines in their MIR spectrum.  The broad silicon features at $\approx$10 $\mu$m is
   clearly visible in all spectra. According to \citet{Bressan06},
   these spectra may represent passively evolving ETGs. The thick black line
   represents  the average  passive ETG spectrum shown in Fig.~\ref{Fig2}. 
   The fluxes have been arbitrary scaled  in order to separate  the spectra.
  }
\label{Fig4}%
\end{figure*}
%

   \begin{figure*}
 \centering
 \includegraphics[width=12.5cm]{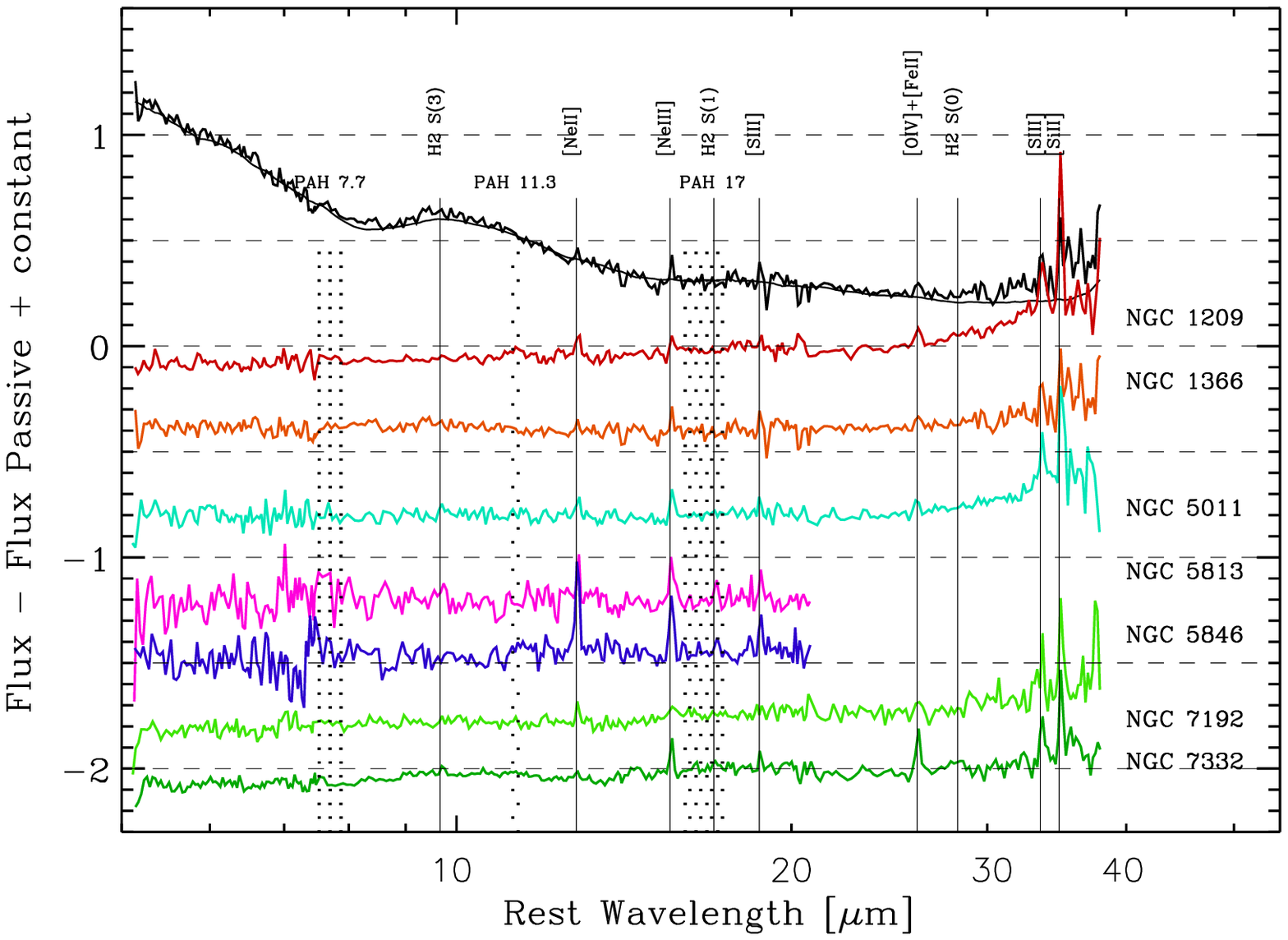}
   \caption{Residuals after an average passive ETG
   spectrum  (the thick black profile, see also Fig.~\ref{Fig2}) has been
   subtracted. As an example,  on the top of the average passive 
   spectrum, we plot the NGC~1366 spectrum.
 We collect {\it Class--1} ETGs, with   ionic and molecular
emission  lines but without clear PAH emission.
Residual fluxes, arbitrarily scaled, confirm  the lack of PAH features. Emission
lines detected are indicated on the top of the plot. 
The NGC~4696 spectrum shows  H$_2$   S(1) and S(3) rotational lines.
 }
 \label{Fig5}%
    \end{figure*}
%

\begin{figure*}
\centering
\includegraphics[width=12.5cm]{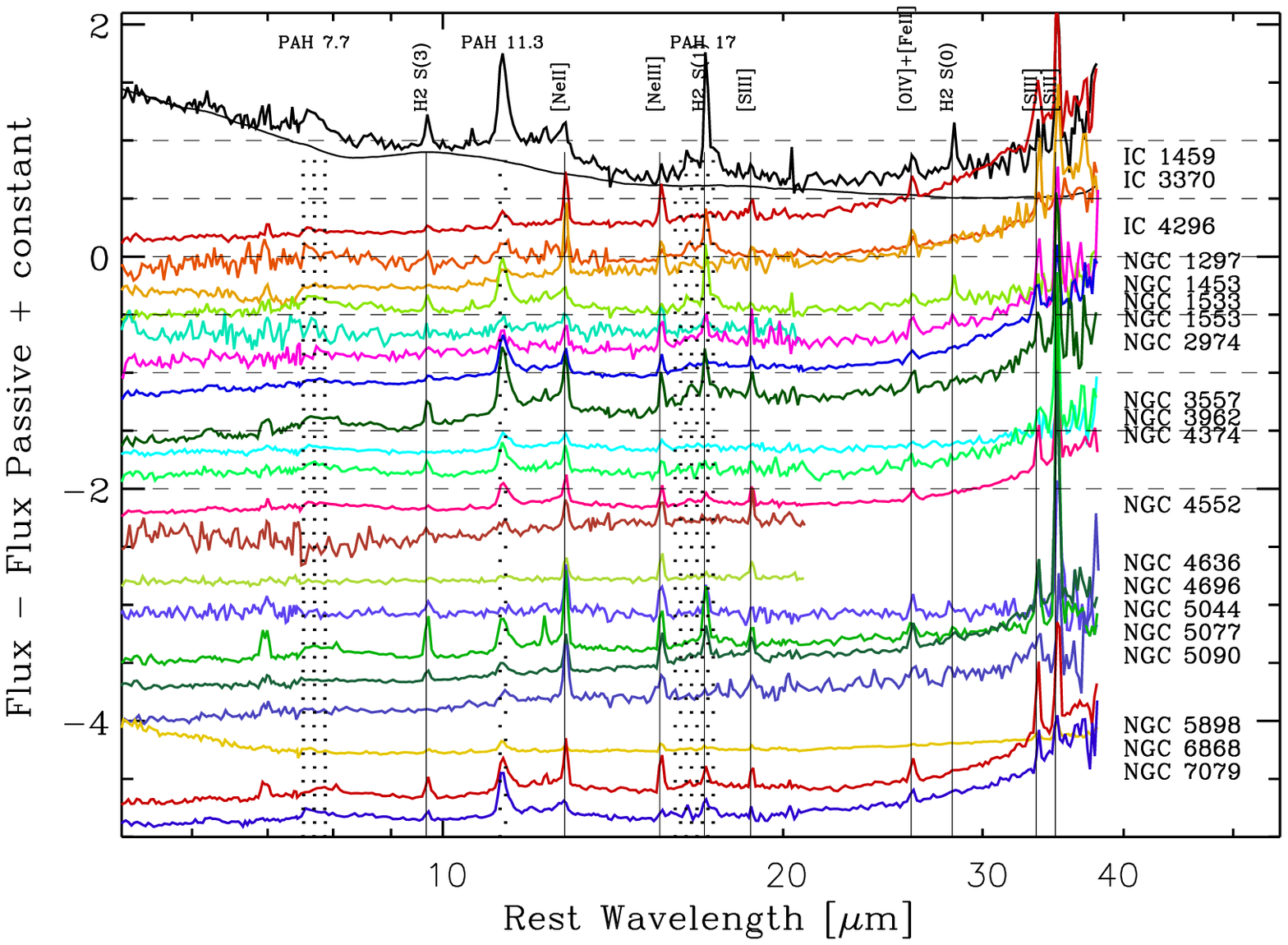}
\caption{{\it Class--2} ETGs, i.e. those
showing prominent ionic emission   lines 
and "unusual" PAH emission, mainly dominated by the 11.3 $\mu$m complex.   
We show the residuals after the subtraction of the average passive ETG spectrum 
(thick black line) as in Fig.~\ref{Fig5}. In some galaxies the broad 10 $\mu$m silicon emission
is still visible (see Fig.~\ref{Fig1}). At the top of the figure we superpose the
spectrum of NGC~1297 on the passive template. 
H$_2$ molecular rotational lines  are detected in some objects.  
Fluxes have been arbitrary scaled  in order to separate  the spectra.}
\label{Fig6}%
\end{figure*}
%

\begin{figure*}
\centering
\includegraphics[width=12.5cm]{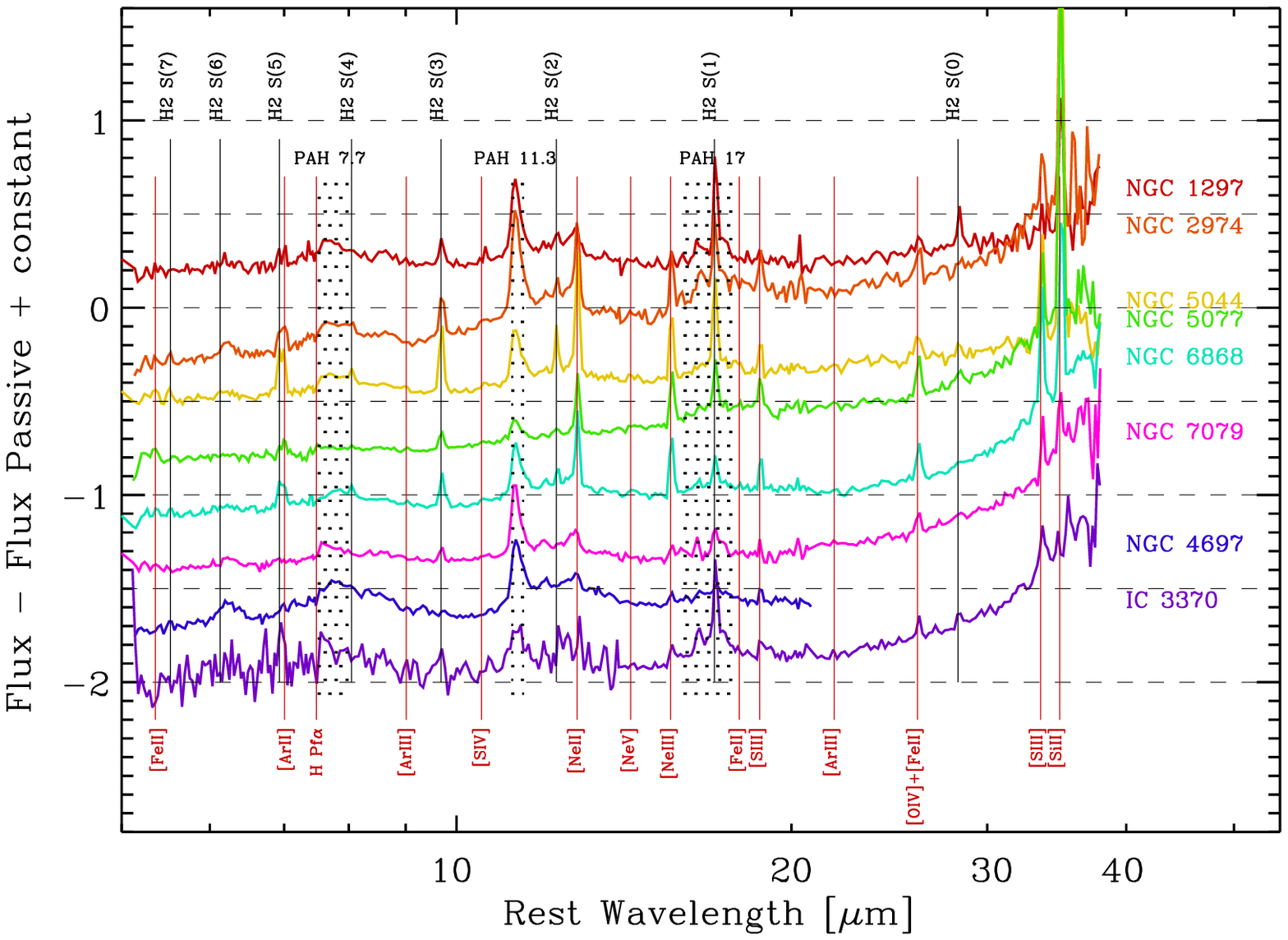}
\caption{{\it Class--2} ETGs.
The MIR spectra of these galaxies display strong H$_2$ molecular 
gas emission. The fluxes have been arbitrary scaled  in order to separate  
the spectra.}
\label{Fig7}%
\end{figure*}
%

\begin{figure*}
\centering
\includegraphics[width=12.5cm]{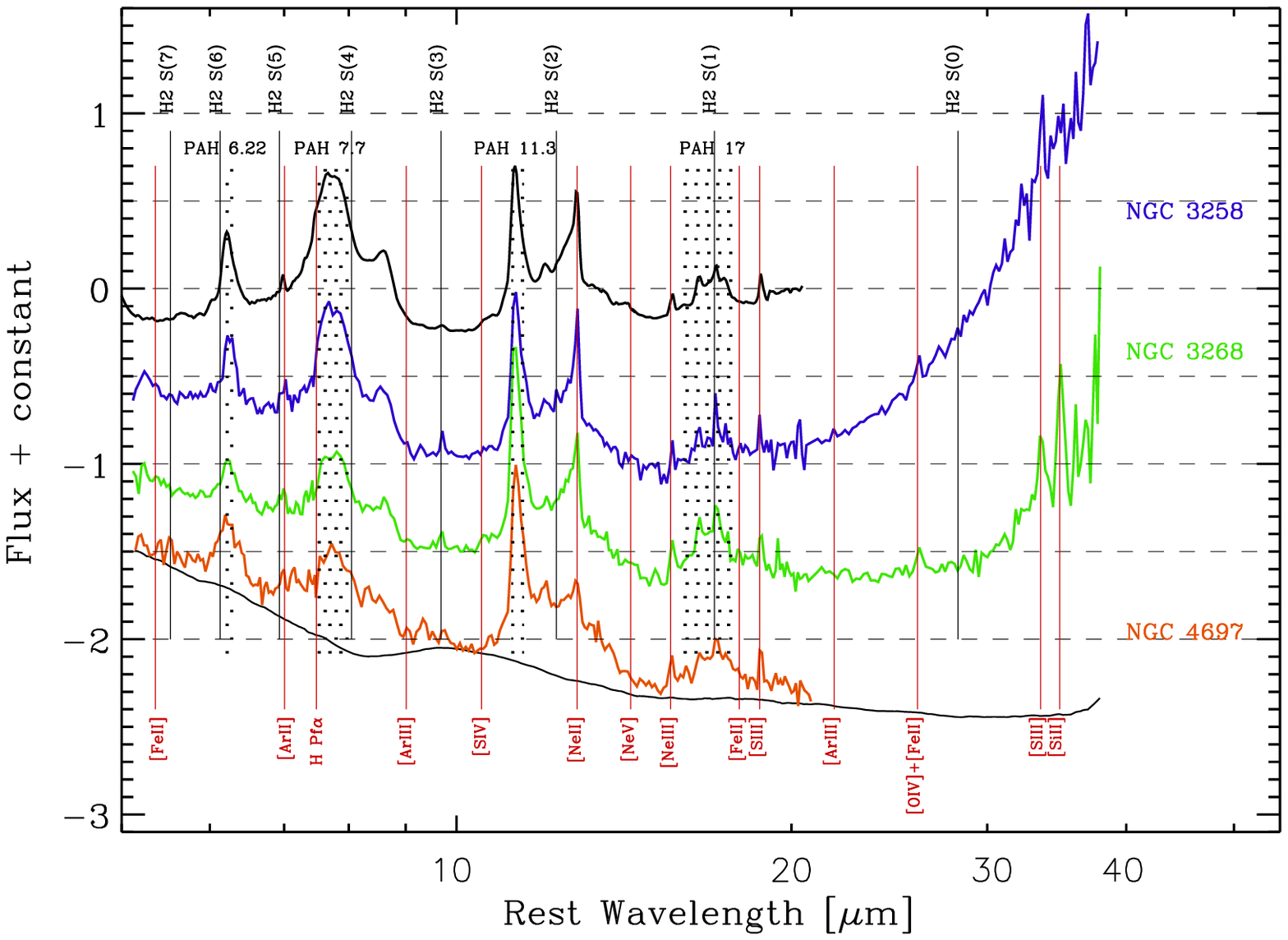}
\caption{ETGs showing ``normal'' PAH features - {\it Class--3}. For comparison, 
at the top of the plot we show the spectrum of the fading starburst NGC~4435 
\citep[see][]{Bressan06,Panuzzo07} and at the bottom the passive 
template. Fluxes have been   arbitrary scaled  in order to separate  the spectra.}
\label{Fig8}%
\end{figure*}


\begin{figure*}
\centering
\includegraphics[width=13.5cm]{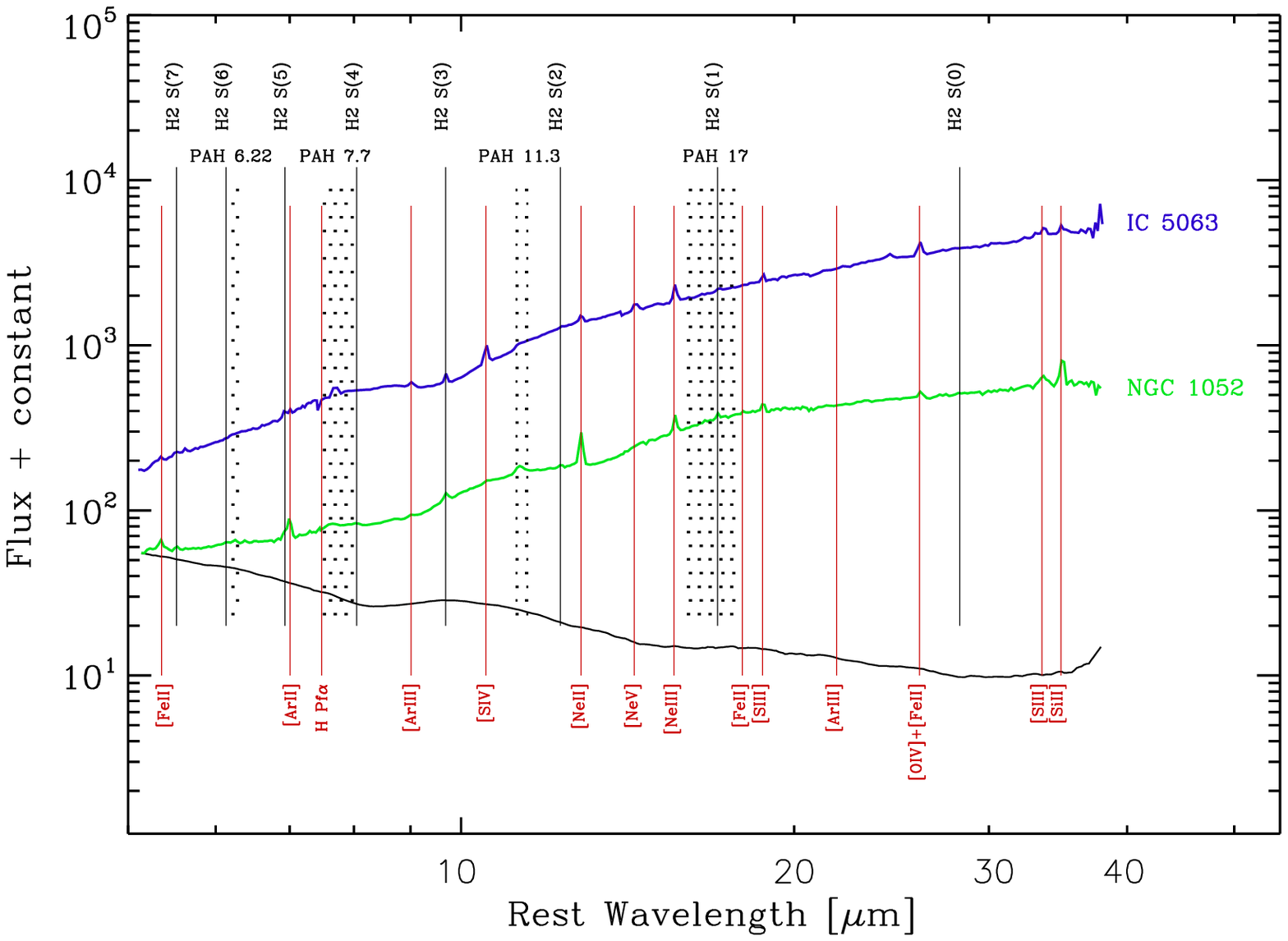}
\caption{ETGs with a hot dust continuum.
We classify such MIR ETG spectra as {\it Class--4}.   H$_2$ molecular lines, 
PAH  7.6 $\mu$m, 11.3 $\mu$m, 12.7 $\mu$m and 17 $\mu$m   complexes and 
ionic line positions are indicated at the top of the spectrum. NGC~1052 is a
prototypical LINER while IC~5063 is an ETG with a Seyfert nucleus.   }
\label{Fig9}%
\end{figure*}
%


\section{The MIR Ne-S-Si diagnostic diagram}
\label{sec:diagnostic}

\begin{figure}
\includegraphics[width=0.48\textwidth]{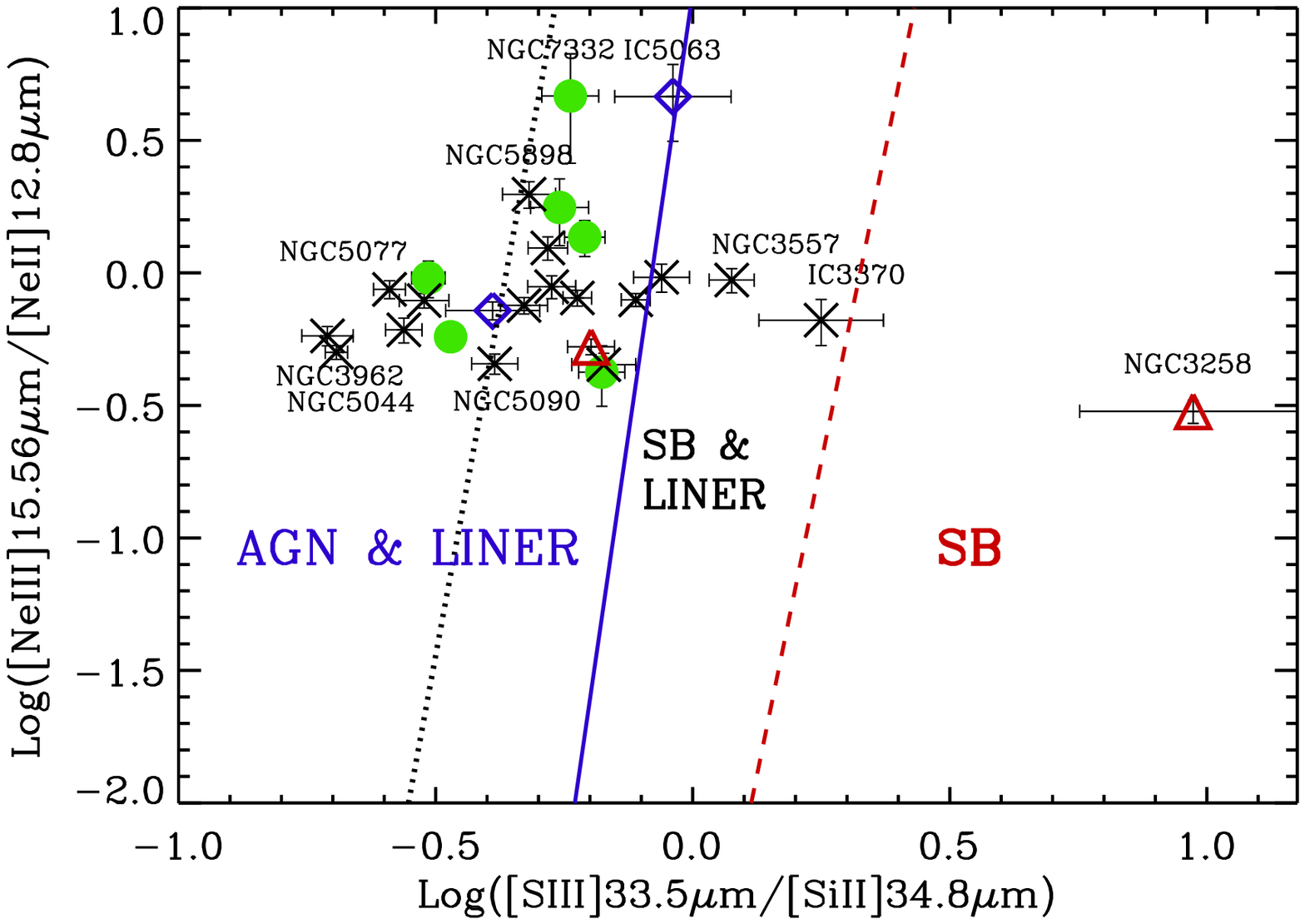}
\includegraphics[width=0.49\textwidth]{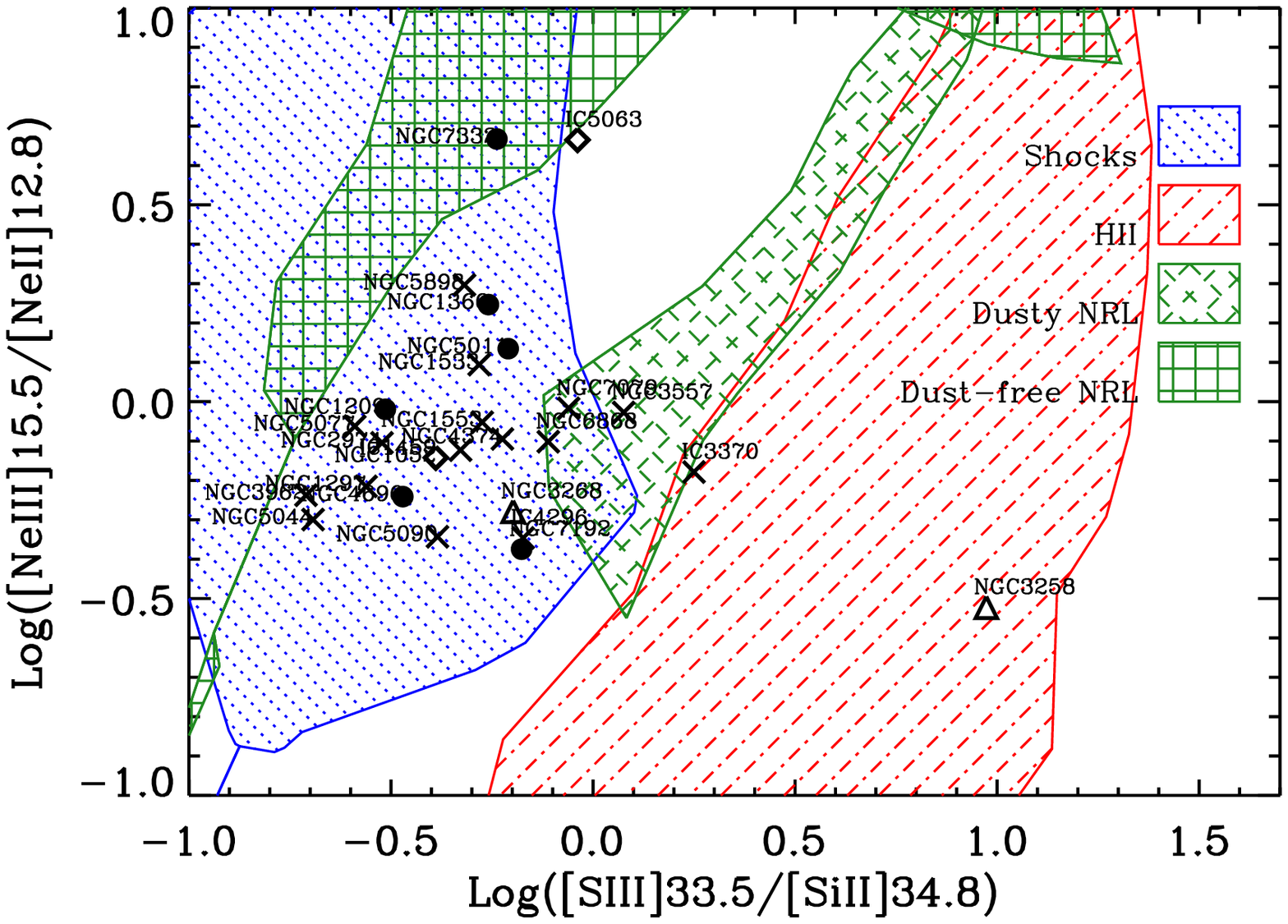}
\caption{
\textit{Upper panel:} Neon, sulphur and silicon diagnostic diagram from \citet{Dale06}.
Dotted, solid and dashed lines divide the plane according to \citet{Dale06} classes: AGN and LINERs 
(classes I and II), HII nuclei and LINERS (class III) and HII regions (class IV).
The different symbols in the diagrams represent the different 
MIR spectral classes of ETG described in Sect.~\ref{sec:analysis} and reported in Table~5:  
Class--1 = filled circles; Class--2 = crosses; Class--3 = triangles; Class--4 = diamonds.
See text for details.
\textit{Bottom panel} \citet{Dale06} diagnostic diagram, i.e.
Log([\ion{Ne}{III}]15.5 $\mu$m/[\ion{Ne}{II}]12.8 $\mu$m) 
vs. the Log([\ion{S}{III}]33.48 $\mu$m/[\ion{Si}{II}]34.82 $\mu$m)  plane.
Models for starburst galaxies \citep[SB:][]{Dopita06}, 
for AGNs \citep{Groves04a,Groves04b}, and for fast shocks
\citep{Allen08} are shown. }
\label{Fig10}
\end{figure}

Nebular emission lines provide a powerful diagnostic tool to
determine the nature of the emission-powering source. 
Classical  optical diagnostic diagrams \citep[\protect{[NII]$\lambda$6583/H$\alpha$
vs. [\ion{O}{III}]$\lambda$5007/H$\beta$, 
[\ion{S}{II}]$\lambda$6716+$\lambda$6731)/H$\alpha$ vs [\ion{O}{III}]$\lambda$5007/H$\beta$, 
and [OI$\lambda$6300/H$\alpha$ vs [\ion{O}{III}]$\lambda$5007/H$\beta$,} 
see e.g.,][]{Baldwin81,Veilleux87}
are used to separate star-formation, Seyferts and LINERs.
In \citetalias{Annibali10a} we showed that the majority of the galaxies in our 
ETG sample are classified as LINERs (see Col.2 of Table~5 for the optical classification).
However, LINERs are compatible with both low accretion-rate  AGN and 
fast shock models in optical diagnostic diagrams, implying that a disentanglement 
of the excitation mechanism is not possible.

The advent of {\it ISO} first, and {\it Spitzer} later, 
allowed astronomers \citep[e.g.][]{Genzel98,Laurent00,Verma03,Peeters04,Armus04,Sturm06,Dale06,Smith07,Dale09}
to investigate diagnostic diagrams in a spectral region with
much smaller dust attenuation and almost free of stellar
absorption features.
Several diagnostic diagrams, based on MIR 
emission lines and/or PAH features, have been proposed to 
characterize the physical state of the system and to infer the nature of the 
dominant powering mechanisms.

The strength of PAH features depends on a complex combination of 
several parameters characterizing the ISM: the metallicity, 
the distribution of sizes and ionization states of 
the PAH molecules, and the intensity and hardness of the 
interstellar radiation field \citep[e.g.][]{Cesarsky96,Li01,Draine07,Galliano08}.  
 
Forbidden lines, like [\ion{Ne}{II}]12.81$\mu$m, [\ion{Ne}{III}]15.55$\mu$m, 
[\ion{S}{III}]18.70$\mu$m, [\ion{S}{III}]33.48$\mu$m, 
and [\ion{Si}{II}]34.82$\mu$m are quite prominent in the MIR spectra 
of star-forming regions, while [\ion{O}{IV}]25.89$\mu$m,
[\ion{S}{IV}]10.51$\mu$m and [\ion{Ne}{V}]14.32$\mu$m are characteristic of  AGN emission
\citep[see e.g.][]{Netzer07,Tommasin08,Tommasin10}.
Bright [\ion{Fe}{II}]25.98$\mu$m and [\ion{Si}{II}]34.82$\mu$m lines are emitted by regions
with different physical conditions like X-ray dominated regions (XDR)
\citep[e.g.][]{Hollenbach99,Kaufman06} 
and shocked regions, where heavy elements such as Si and Fe may
be less depleted from the gas phase due to dust 
destruction \citep[e.g.][]{OHalloran06}.

In the top panel of Fig.~\ref{Fig10}, we report the position of our
ETGs in the diagnostic diagram 
[\ion{Ne}{III}]15.56$\mu$m/[\ion{Ne}{II}]12.8$\mu$m vs. [\ion{S}{III}]33.48$\mu$m/[\ion{Si}{II}]34.82$\mu$m
proposed by \citet{Dale06}. The [\ion{Ne}{III}]15.56$\mu$m/[\ion{Ne}{II}]12.8$\mu$m ratio 
represents the hardness of the ionizing source, while the 
[\ion{S}{III}]33.48$\mu$m/[\ion{Si}{II}]34.82$\mu$m is affected by the 
presence of X-rays and Si depletion.

The diagonal lines in the plot show empirical separation of different 
powering sources by \citet{Dale06}. 

In the bottom panel of Fig.~\ref{Fig10}, we report models for HII regions,
Narrow Line Regions (NLRs) and shocks,
collected with the help of the {\tt ITERA} software\footnote{\tt 
http://www.strw.leidenuniv.nl/$\sim$brent/itera.html} (Groves \& Allen, in prep). 
The model libraries are from \citet{Dopita06} (HII regions),  
\citet{Groves04a,Groves04b} (NLR) and \citet{Allen08} (shock and shock+precursor).
Two sets of Narrow Line Region models were used, one dust-free, the 
second accounting for the presence of dust in the NLR, including the effect of metal depletion. 

We can note that HII regions powered by star-formation  have a very small
overlap with NLR models in this diagnostic diagram, while dust-free NRL models
almost completely overlap with shock models.

Since the metal depletion in NLR is not known and most probably
varies between different sources \citep{Dopita02}, we can expect the region 
of the above diagnostic diagram between dust-free and dusty NRL models to be
 populated by NRLs with different dust-to-gas mass ratios. As a consequence,
sources in the region with Log([\ion{S}{III}]33.48$\mu$m/[\ion{Si}{II}]34.82$\mu$m)
between -1 and 0 can be explained both as shock-powered and AGN-powered.
Most of our  ``active'' sample lie in this region, so it is not possible to distinguish if the
line emission is powered by shocks or by AGN  using this
diagnostic diagram. We finally note that only NGC~3258 can be classified  
as a star-forming object from its emission lines.


\section{Discussion}
\label{sec:discussion}

\begin{figure}
\centering
\includegraphics[width=0.45\textwidth]{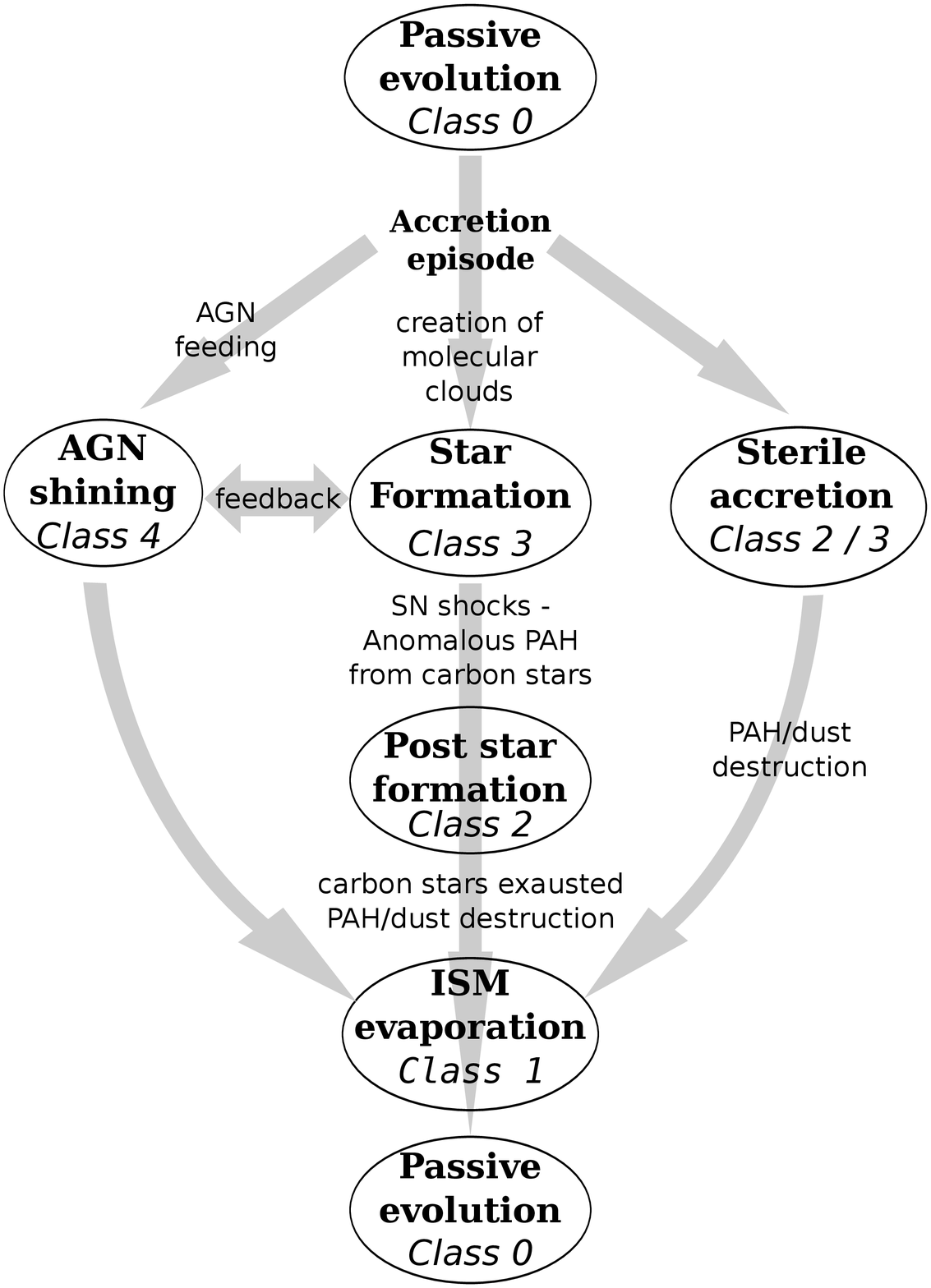}
\caption{Schematic representation of the evolutionary path triggered by a 
perturbation/accretion event. The labels in bold show the physical phases, while 
labels in italics show the associated MIR class; arrows show the
paths that an ETG can follow.}
\label{Fig11}
\end{figure}

Several kinds of signature suggest that ETGs in our sample have
undergone through renjuvenation episodes in their recent history. 
Our analysis of the optical line-strength indices \citepalias{Annibali07}
shows that ETGs  have different luminosity weighted ages 
(see column 7 in Table 1). This is widely believed 
to be due to recent  secondary star formation events 
induced by accretion and/or interaction.
In Table A1 of \citetalias{Annibali10a} we report kinematical and 
morphological peculiarities, as they appear in the current literature, 
of our ETGs. A large fraction
show kinematic, like gas vs. star counter-rotations, 
and morphological distortions, like tails and shells, 
supporting the view of accretion events. In particular, shell-like 
structures support the view of minor accretion events 
\citep{Dupraz86,Ebrova09}.
In \citetalias{Annibali10a} we show that the nebular oxygen
abundance is slightly lower than that of stars suggesting an
external origin for at least part of the gas present in the nuclei
of our ETGs. 

In  the {\it GALEX} study of our ETGs \citepalias{Marino10}
we also show that, for a small fraction of barred lenticulars, the
rejuvenation of the nucleus could also be attributed to
inner secular evolution likely driven by bars.
At the same time, accretion of gas from the outer 
halo of an ETG, induced by interaction, could also be possible.
Examining 33 galaxies representative of the SAURON sample
\citet{Oosterloo10} detected HI in 2/3 of their field
sample, concluding that neutral gas accretion
plays a role in the evolution of field ETGs.
In a survey of about 100 ATLAS3D galaxies, \citet{Serra09} 
detected HI in about half of the ETGs located outside Virgo. 
They suggest that isolated ETGs are characterized 
by very regular, rotating cold gas systems while 
galaxies with neighbours or residing in 
galaxy groups typically have disturbed HI morphology/kinematics. 
They conclude that environment is a fundamental driver of 
ETG evolution at z$\sim$0 and that, in the absence of significant disturbance 
from nearby objects, ETGs can accrete and maintain large systems of 
cold gas without this appreciably affecting their optical morphology.
From the above studies, it appears ``normal'' that ETGs in low density 
environments may acquire gas and the material associated with this gas 
\citep[as beautifully shown by dust-lanes in NGC 1409/1410 described in][]{Keel04}
during perturbation events. This material may lose angular momentum falling
towards the galaxy centre and may or may not feed star formation
or AGN activity.

Finally, we cannot exclude ``dry'' accretions, i.e. accretion of pure stellar systems.
The spectral analysis of faint companions of ETGs in low density 
environments, where our ETGs are mostly located, performed 
by \citet{Grutzbauch09} showed that the 
vast majority of these latter do not show emission lines
\citep[see also the population analysis in][]{Annibali10b}.

Given the great diversity of the {\it nuclear} MIR spectra of galaxies presented here,
it is natural to ask if the above empirical spectral classes
represent different phases of a unique ``evolutionary path''
driven by accretion, secular evolution and interaction events which re-fuel
the nucleus.
We sketch an evolutionary path in Fig. \ref{Fig11}.
A passive galaxy, under the perturbation exerted by an external
accretion, interaction or a secular evolutionary mechanism, could build a reserve of dense
ISM in the nucleus which can switch-on the AGN and/or star-formation activity.
Alternatively, if the accreted gas mass is too small, a sterile accretion
could take place.
The dense gas is then consumed by the AGN and/or star formation, 
or evaporated by the interaction with the hot gas halo; the galaxy would 
eventually return to the passive phase.
We don't treat the case of dry accretion which should not leave
signatures in the MIR spectra.

We now try to associate to each physical phase of this evolutionary path
a characteristic MIR spectral class.

As discussed above, the passive phase is characterizer by Class-0 MIR 
spectra, since this is the only class that doesn't show any emission lines. 
We detected only the silicon features at about 10 and 18 $\mu$m, arising from
the circumstellar envelopes of O-rich AGB stars. This dust, once diffused in
the ISM, it is destroyed by the hot gas halo \citep[see][]{Clemens10}.

The star-formation phase is characterized by spectra of Class 3. This class is 
also characteristic of the phase just after the switch-off of the star formation episode 
when the UV flux from B stars is still abundant enough to create ionized PAHs, the 
principal carriers of the 7.7$\mu$m feature.

The Class-4 spectra, with their hot dust continuum, are characteristic of 
the AGN phase. However we note that AGN activity is frequently
associated with star-formation activity \citepalias[see also][]{Annibali10a} so
Seyfert galaxies will be characterized by spectra mixing Class-4 and Class-3
\citep[see][for a collection of MIR spectra of Seyfert galaxies]{Wu09}.
Note that in some cases the AGN MIR spectra can be dominated by
synchrotron emission, like in M87 \citep{Buson09}.

The origin of Class-2 spectra is discussed in detail in \citet{Vega10}, and 
they can be associated with the post-star-formation phase. In this phase, there are three
mechanisms that concur, resulting in unusual PAH features:
i) the UV radiation field is no longer strong enough to ionize PAH molecules
ii) shocks from supernovae destroy small PAHs; iii)
the production of PAHs from carbon stars.
We note that, for solar metallicity, carbon stars are present in
stellar populations with ages between $\sim$250 Myr and $\sim$1.3 Gyr 
\citep{Marigo07}. If we assume a characteristic life time of 200 Myr for
star formation episodes \citep[see e.g.][]{Panuzzo07}, we espect a ratio between
Class-2 and Class-3 of $\sim$5, which is consistent with the ratio observed in this 
sample ($\sim$7).
Extremely interesting among the MIR class 2 ETGs are the
spectra showing H$_2$ molecular gas emission (see Fig.~\ref{Fig7}).
H$_2$ formation is an important process in post-shock 
regions, since H$_2$ is an active participant in the cooling
and shielding of the region \citep{Cuppen10,Guillard10}.
In our view, although we cannot disentangle using the Dale-plot
in Fig.~\ref{Fig10} between AGN and shock powering mechanisms, 
these latter could play a significant role in the centre of ETGs.

In our evolutionary path we envisage that an accretion 
episode can happen without the 
ignition of an AGN nor of star-formation activity (Sterile accretion in Fig.~\ref{Fig11}).
There is probably no unique MIR spectral class charaterizing this phase. 
The acquired material may contain PAH molecules with different
size distributions, depending on the origin of the gas, and on the
efficiency of the sputtering effects of the hot ISM. The ionization
fraction also depends on the UV radiation fields which depends on if the
accretion was limited to pure gas or if also relatively young stars were
acquired. This phase is probably characterized by MIR spectra of class 2
\citep[with the exception of features associated to carbon stars, as in ][]{Vega10}
or class 3, if the UV radiation field is strong enough.

Finally we envisaged in our evolutionary path an ``ISM evaporation'' phase.
For objects that had star formation activity, this phase follows the 
post-star-formation phase; carbon stars are exhausted (so no more PAH production) 
and the PAHs were destroyed by the hot plasma halo in which ETGs are immersed.
The residual gas is then ionized by either post-AGB stars or by SN Ia shocks.
For the objects that experienced the AGN phase, this is the phase where the
gas reservoir of the AGN is exhausted and the observed low ionization emission lines 
are due to either a residual low power AGN activity or to shocks and/or PAGB stars.
To this phase we associate Class-1 spectra which show only low ionization emission lines.
When all the warm ionized gas is transformed into the hot plasma phase, the
galaxy will return to the passive phase.

A more complete discussion on the evolutionary path described above,
including the use of information from other spectral windows, is
beyond the scope of this article and will be the subject of a future article.
However we showed that, although all ETGs in our sample are quite homogeneous in the
optical window and are mostly classified as LINERs, the MIR spectral region
is very useful in identifying the physical process occurring in these galaxies.


\section{Conclusions}
\label{sec:conclusions}

We presented {\it Spitzer}-IRS spectra for 40 ETGs, 
18 of which from our own proposal (ID 30256), and 22 from the archive.
The entire data-set of 40 galaxies were analyzed homogeneously 
for internal consistency.
The 40 galaxies belong to our original sample of 65 ETGs (the
R05+A06 sample),
for which we had previously characterized the stellar populations 
and the properties of the optical emission lines.
From optical emission line ratios, the majority of our galaxies are 
classified as LINERs.

We measured the MIR atomic/molecular lines and the PAH features 
through an ad-hoc algorithm that includes an adequate treatment 
of the emission from the old stellar population, whose contribution 
to the MIR spectra of ETGs cannot be neglected.
The template of the old stellar population was obtained as the 
average of the MIR fluxes of the three passive ETGs with the 
highest signal-to-noise ratios  in our IRS spectra.

We propose a new classification of ETGs based on the MIR 
properties of the continuum, the atomic and molecular lines, 
and the PAH complexes:

\begin{itemize}

\item Class--0 comprises {\it passively evolving ETGs}. These galaxies are characterized by :
(a) the absence of PAH, atomic and molecular line emission; (b) a dip at 8~$\mu$m probably due to
photospheric SiO absorption bands; c) a broad emission feature around 10~$\mu$m due to 
silicate emission bands arising from the dusty circumstellar envelopes of O-rich AGB stars;
d) a broad emission feature around 10~$\mu$m with the same origin.
This class accounts for $\sim$ 20\% of our sample.

\item Class--1 includes ETGs that exhibit atomic and molecular emission lines, but that still
do not show PAH features. Class--1 accounts for $\sim$ 17.5\% of our sample.

\item Class--2 includes ETGs which exhibit PAH features (besides atomic and molecular emission lines) 
with {\it unusual} interband ratios: usually strong emission features at 6.2, 7.7, 
and 8.6~$\mu$m are weak in contrast to prominent features at 12.7, 11.3 and 17~$\mu$m.
Indeed, a result from {\it Spitzer} is that this is more the rule than an exception in ETGs.
Class--2 accounts for $\sim$ 50\% of our sample.

\item Class--3 comprises ETGs  with {\it normal} PAH interband ratios, i.e. more typical of 
star-forming galaxies. {\it Class--3} represents $\sim$7.5\% of the sample.

\item Class--4 includes ETGs  with hot dust continuum.
AGN [\ion{S}{IV}] and [\ion{Ne}{V}] high-ionization emission lines are revealed
only in IC 5063. Class--4 represents $\sim$5\% of our sample.

\end{itemize}

Based on starburst, AGN, and shock models, we investigated MIR line-ratio
planes to spot diagnostics that allow 
for a complete disentanglement of the different excitation mechanisms. 
We used the diagram [\ion{Ne}{III}]15.55$\mu$m/[\ion{Ne}{II}]12.8$\mu$m vs. 
[\ion{S}{III}]33.48$\mu$m/[\ion{Si}{II}]34.82$\mu$m
to attempt a disentanglement of these mechanisms in the galaxy nuclei.
The diagram has the advantage that it involves strong and easy to measure 
emission features. [\ion{S}{III}]33.48$\mu$m and [\ion{Si}{II}]34.82$\mu$m lie in the 
same Long-High module of the {\it Spitzer}-IRS,  which minimizes cross 
module uncertainties involving calibration and aperture matching.
When comparing our data with this diagnostic, it emerges that 
the majority of ETGs are outside the starburst area and
lie in the region in which either shock, 
AGN-like, or both mechanisms give rise to the observed 
emission. 

We sketched a possible evolutionary scenario for ETGs induced by accretion 
and interaction events. Such episodes are indeed suggested both by the
kinematical/morphological peculiarities and by the nebular O
abundance found in \citetalias{Annibali10a} of the series. 
We associated to the different phases of the above scenario the
MIR spectral classes defined above. This association and the variety of MIR
characteristics allow to shed light
on the renjuvenation episodes and on the physical processes happening
in ETGs (in particular in LINERs), which are, otherwise, quite
homogeneous in the optical spectral window.

Further insight will come from the study of 
the PAH emission and of the H$_2$ molecular lines
\citep[see however][]{Vega10}, and from the comparison of MIR 
spectral characteristics with the information from other 
wavelenghts, which will be investigated and discussed in forthcoming papers.

\begin{acknowledgements}
This work is based on observations made with the Spitzer Space Telescope, 
which is operated by the Jet Propulsion Laboratory, California Institute of 
Technology under a contract with NASA. Support for this work was provided by 
NASA through an award issued by JPL/Caltech.
Part of this work was supported by ASI-INAF contract
I/016/07/0. We thank the referee for the very useful suggestion.
RR is highly indebted for the hospitality and financial
support of the DSM/Irfu/Service d'Astrophysique, CEA Saclay, France.
\end{acknowledgements}


\end{document}